\documentclass[9pt]{elife} 

\usepackage{lipsum} 
\usepackage[version=4]{mhchem}
\usepackage{siunitx}
\DeclareSIUnit\Molar{M}
\usepackage{textgreek}

\title{Statistics of \textit{C.~elegans} turning behavior reveals optimality under biasing constraints}

\author[1,\authfn{1}]{W. Mathijs Rozemuller}
\author[1,2,\authfn{2}]{Steffen Werner}
\author[3,\authfn{3}]{Antonio Carlos Costa}
\author[3]{Liam O’Shaughnessy}
\author[3,4]{Greg J. Stephens}
\author[1*]{Thomas S. Shimizu}

\affil[1]{AMOLF Institute, Amsterdam, The Netherlands}
\affil[2]{Department of Experimental Zoology, Wageningen University, Wageningen, The Netherlands}
\affil[3]{Department of Physics and Astronomy, Vrije Universiteit, Amsterdam, The Netherlands}
\affil[4]{Okinawa Institute of Science and Technology, Onna-son, Okinawa, Japan}

\corr{shimizu@amolf.nl}{TSS}

\contrib[\authfn{1},\authfn{2}]{These authors contributed equally to this work}

\presentadd[\authfn{3}]{Laboratoire de Physique de l’Ecole normale sup\'erieure, Paris, France}

\begin{document}

\maketitle

\begin{abstract} 
Animal locomotion is often subject to constraints arising from anatomical/physiological asymmetries. We use the nematode  \textit{C.~elegans} as a minimal model system to ask whether such constraints might shape locomotion patterns optimized during evolution. We focus on turning behaviours in two contrasting environmental contexts: (i) random exploration in the absence of strong stimuli and (ii) acute avoidance (escape) navigation upon encountering a strong aversive stimulus. We characterise the full repertoire of reorientation behaviours, including gradual reorientations and various posturally distinct sharp turns. During exploration, our measurements and theoretical modelling indicate that orientation fluctuations on short timescales are, on average, optimized to compensate the constraining gradual turn bias on long timescales. During escape, our data suggests that the reorientation is controlled not by an analog logic of continuous turn-amplitude modulations, but rather through the digital logic of selecting discrete turn types, leading to a symmetric escape performance despite strongly asymmetric turning biases.
\end{abstract}

\section{Introduction}

Changing course effectively is fundamental to exploration and navigation by motile organisms. Although the mechanisms that underlie motility are diverse, the motile strategy of many organisms can be described as a biased random walk, in which the statistics of reorientation (turns) determine how the organism explores space, and are modulated to achieve navigation \cite{Berg1993,Codling2008}. An ubiquitous yet underappreciated aspect of turning, as implemented in these strategies, is the need to overcome inherent asymmetries in the actuating anatomy and control physiology to optimise performance. The biased random walk is effective as a motile strategy precisely because the bias is imposed in response to the environment. Biases arising from internal factors uncorrelated with the environment can thus generically be expected to diminish performance and in turn act as constraints in optimizing behavior.

One simple strategy for overcoming such biasing constraints is to randomise the direction of the bias relative to the environment. This is implemented, for example, in the swimming kinematics of \textit{E. coli} bacteria, which effectively eliminate the effects of anatomical asymmetry (due e.g. to the random positioning of flagella on the cell surface) by incessantly rolling their cell body about the direction of motion as they explore their environment by swimming (runs), thereby eliminating correlations between any turning bias and the environment \cite{Berg2004}. As a result, the bacterium can optimise its exploration and navigation performance by tuning only two variables --- the temporal frequency at which erratic turns (tumbles) are generated, and the average (unsigned) angle of tumble-induced turns \cite{Berg1972}.

By contrast, within the animal kingdom, the kinematics of locomotion tend to more faithfully reflect the asymmetries of animal anatomy. The vast majority ($>99\%$) of animal species have a bilateral body plan, meaning radial symmetry about the anterior-posterior (A-P) axis is broken to yield two more orthogonal axes: the dorso-ventral (D-V) and left-right (L-R) axes \cite{Levin2005}.
This three-axis anatomical frame has tangible advantages for directed locomotion as it enables distinct motor patterns in the vertical (\textit{i.e.}~gravitational) and lateral directions, across which resource distributions tend to be very different \cite{Hollo2012}. But these advantages also come with a cost: breaking rotational symmetry eliminates the possibility of removing detrimental directional bias by rotation about the locomotion direction. Indeed, the statistics of turning behaviour demonstrate some degree of lateral bias (handedness) across the immense diversity of bilaterian animals, from invertebrates \cite{Gray2005,Peliti2013,Buchanan2015,Ayroles2015,Helms2019} to humans \cite{Souman2009,Bestaven2012}.

The causal factors that underlie turning biases are not well understood, but are likely to be diverse in origin, given the broad spectrum of reported phenomenologies. Some instances of reported turning bias were in the gradual reorientation (path curvature) during intervals of relatively smooth forward or backward motion \cite{Souman2009,Peliti2013,Helms2019}. In other examples, turning bias was observed in the more discrete sharp-turn behaviours that occur either spontaneously during unbounded locomotion \cite{Gray2005} or upon forced decisions in a Y-maze \cite{Buchanan2015,Ayroles2015}. Some of these turning biases persisted throughout the lifetime of an individual \cite{Gray2005,Buchanan2015,Ayroles2015}, whereas at least one example of a gradual turning bias was found to vary, and even change sign, over relatively short time scales \cite{Souman2009}. The mechanistic origins are therefore likely different from case to case, and relatively few studies have been able to directly address causal factors experimentally. Nevertheless, the available data point towards neural control physiology \cite{Gray2005,Buchanan2015,Bestaven2012}, rather than hard constraints at the anatomical level, as causative factors for turning bias during locomotion.

In this study, we investigate bias in the turning behaviour of the nematode \textit{C. elegans}, arguably the simplest and best characterised animal model for locomotion \cite{Gjorgjieva2014}. The adult hermaphrodite body plan comprises just $959$ somatic cells \cite{White1988}, of which $302$ are neurons with a fully mapped connectome \cite{White1986,Cook2019, witvliet2021connectomes}. Despite this compact anatomy, these worms perform a variety of locomotion tasks, such as exploration for food, chemotaxis, and escape. The relative simplicity of its anatomy, control physiology, and postural kinematic repertoire positions \textit{C. elegans} as a compelling minimal model system to address fundamental questions about behavioural strategies of animal locomotion and the underlying neural control mechanisms. Previous work has identified significant orientational biases in \textit{C. elegans} turning behavior \cite{Gray2005,Peliti2013,Helms2019}, but how such biases affect performance in specific locomotion tasks remains an open question.

Turning during \textit{C.~elegans} locomotion occurs both abruptly and gradually as the worm crawls on surfaces.
\textit{C.~elegans} crawls while lying 'on its side', with its L-R body axis normal to the surface.
Crawling is driven by undulatory propulsion, in which body bends in the D-V direction are initiated near the head and propagated along the length of its anatomy towards the tail, resulting in postures and trajectories of motion that are approximately sinusoidal in shape. Occasionally, these forward runs are interrupted by sharp reorientation events in which the body bends deeply to generate a large change in orientation\cite{Croll1975}, as well as brief reversal events in which propagation direction of the body wave (and hence also the worm's movement direction) is inverted but do not otherwise change the direction of motion. When navigating environmental gradients by chemotaxis, reversals and sharp turn events are often clustered in time, generating short intervals of frequent turning (pirouettes) that interrupt otherwise smooth crawling trajectories. Modulating the temporal frequency of pirouettes in response to the environment provides \textit{C. elegans} with one mechanism for biasing its random walk to achieve chemotaxis \cite{Pierce1999}. Between these abrupt reorientation events, more gradual changes in orientation also occur, resulting in trajectories that are curved and meandering over length scales greater than those of the aforementioned body wave \cite{Peliti2013,Helms2019}, and it has been shown that worms can also bias this curvature in response to environmental gradients to enhance chemotactic performance \cite{Iino2009,Luo2014_Cell}.

Directional biases are known to exist for both sharp and gradual turns of \textit{C. elegans} \cite{Peliti2013,Broekmans2016}, even in the absence of environmental gradients, yet their impact on locomotion performance has yet to be studied systematically. The sharp turns are known to be strongly biased in the ventral direction \cite{Croll1975,Gray2005}. The biological reason for this D-V bias is not well understood, but it evidently reflects neural control, as ablating a single neuron (RIV) results in a much larger fraction of sharp turns in the dorsal direction, largely eliminating this bias \cite{Gray2005}. Bias in gradual turns is less well characterized, perhaps because it is an inherently long time- and length-scale phenomenon, and thus accurately quantifying it requires extensive statistics from long trajectory recordings. However, recent studies have provided evidence that orientational statistics of crawling \textit{C. elegans} trajectories collected over $30$-$80$ min are not isotropic \cite{Peliti2013}, and that biases in gradual turns (of $\sim$2 degrees per second) can be detected even within $30$ min trajectories \cite{Helms2019}.

Here, we present an extensive study of \textit{C. elegans} turning statistics in two contrasting behavioural contexts: exploration in the absence of environmental gradients, and escape upon encountering a strong aversive stimulus. To sufficiently sample both sharp- and gradual-turn statistics within each individual, we recorded long ($120$ min) trajectories within an arena enclosed by a repellent boundary. The worms spend most of their time exploring the arena by freely crawling in the absence of environmental gradients, allowing us to sample the statistics of "spontaneous" turns, which the worms execute without a triggering environmental stimulus. In addition, upon encountering the boundary impregnated with a chemical repellent, an escape response \cite{Ghosh2012,Mohammadi2013,Leung2016} is triggered, causing the worm to turn around. We quantify the strength of orientational biases in both sharp and gradual turns, and quantify their effects on performance in both behavioural contexts.
This reveals how biases in turning behavior impacts both exploration and navigation, and that optimization of turn statistics can be leveraged to increase performance despite anatomical or physiological constraints.

\section{Results}

\subsection{Worms in repellent-boundary arenas demonstrate both exploration and escape behaviors}

Globally, the worm's motile behaviour can be described as intervals of relatively straight forward motion (runs) interrupted by brief intervals of backward motion (reversals) and abrupt reorientation events (sharp turns). Some gradual reorientation also occurs during runs, which -despite being comparatively subtle- could cumulatively impact the navigational strategy over long times \cite{Kim2011, Helms2019, Stephens2010}. Reversals have a negligible net effect on orientation once the worm reverts to forward motion. Sharp reorientation events are defined as any event where the worm performs a deep body bend and folds onto itself.

To extensively sample the reorientation behaviour of worms, we imaged the motility of \textit{C. elegans} strain N$2$ individuals crawling on agar plates for a duration of approximately $2$ hours (Methods). Worms were confined by a repellent boundary to within a $\SI{38}{\milli\meter} \times \SI{38}{\milli\meter}$ arena, the entirety of which can be imaged at a resolution sufficient to resolve not only the worm's position but also its postural dynamics (effective pixel size $\SI{18.7}{\micro\meter}$). The motility of up to eight worms was measured simultaneously within these arenas. Throughout the duration of the measurement, the identity of each worm was tracked, enabling us to study variability in behaviour between individuals. The arena contained no food, and worms were kept off food for $15$ minutes prior to the experiment, to reduce non-stationarities in behaviour that are known to affect worm motility (including sharp turn rates) for several minutes after transitioning from an on-food to an off-food environment \cite{Gray2005,Salvador2014,Broekmans2016}. Reorientation statistics of $100$ individual worms were obtained by analysing the video recordings (Methods), yielding a total of $12,475$ sharp reorientation events within $197$ hours of trajectories. In $85\%$ of these identified sharp turns, postural dynamics could be unambiguously resolved, thus allowing analysis of postural kinematics for $98\%$ of the total trajectory duration.

\begin{figure}[t]
\centerline{\includegraphics[width=\textwidth]{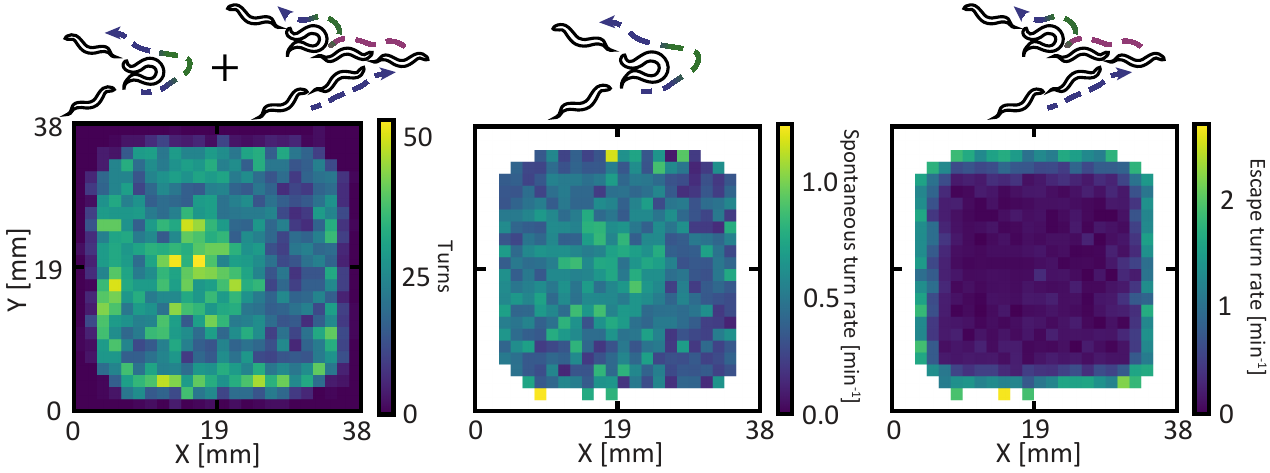}}
\caption{\textbf{Spatial distribution of two types of sharp turns.} (left) The spatial distribution of both spontaneous sharp turns and escape turns inside the arena. (Middle) Spontaneous turns, defined as sharp turns that occur during forward crawling, occur at an equal rate. (right) The escape turn rate, defined as sharp turns following a reversal, is sharply increased near the repellent boundary of the arena. Pixels at the edge that have been occupied by a worm with fewer than $5$k data points ($\approx \SI{7}{\min}$) are not included.}
\label{fig:TurnPartitioning}
\figsupp[Worm collisions minimally impact the trajectory dynamics.]{Worm collisions minimally impact the trajectory dynamics.
(A) The sharp turn frequency is not affected by the collusion events as the total number of sharp turns (summed up over all worms and all times) is approximately the same before and after a collision ($t=0$ marks the midpoint between first and last contact with another worm during a collision).
The dip at $t=0$ is from the duration of the collision.
(B) A collision event has no long-term effects on the speed. Shaded regions show the $95\%$ confidence interval, bootstrapped for collisions.}{\centerline{\includegraphics[width=\textwidth]{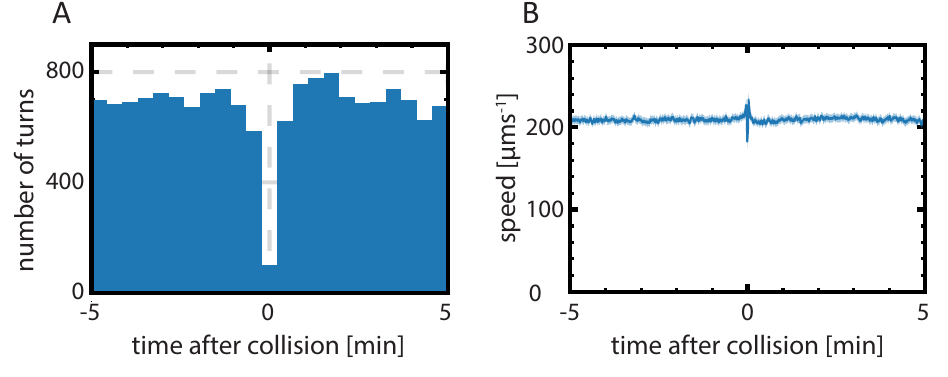}}}
\figsupp[Average speed across all worms remains mostly constant.]{The average speed across all worms has a small increase during the first $\SI{20}{min}$, but remains constant for the remaining duration of the measurement.}{\centerline{\includegraphics[width=\textwidth]{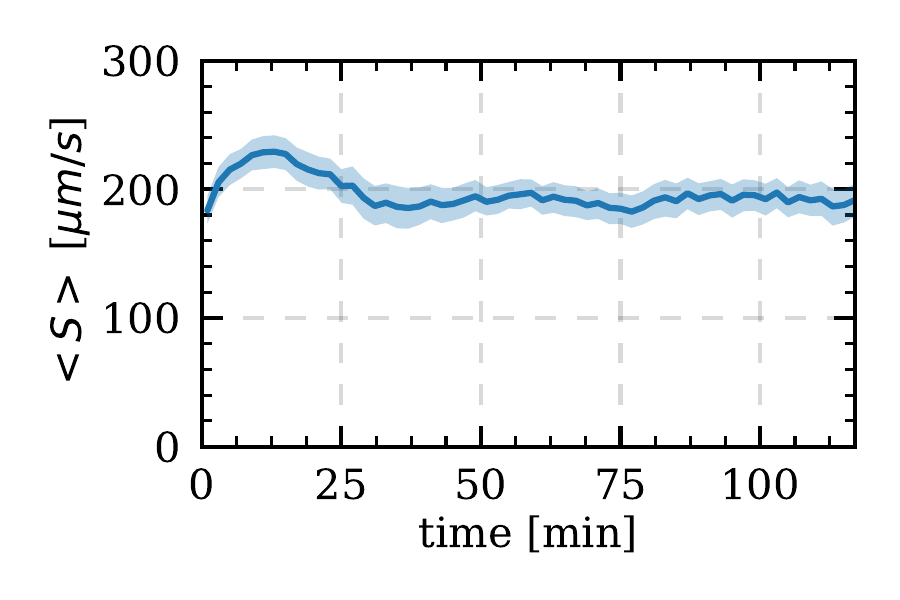}}}
\end{figure}

The repellent boundary not only served to confine worms to within the camera's field of view, but also acts as an aversive stimulus. This allowed us to study reorientation statistics in two contrasting behavioural contexts within a single measurement:
(i) exploration in the absence of environmental stimuli and (ii) the escape response upon encountering the strongly aversive repellent boundary.
For most of the trajectory duration, worms engaged in exploratory behaviour across the arena, characterised by long runs interrupted by "spontaneous" sharp turns \cite{Srivastava2009,Broekmans2016}. Upon encountering the repellent boundary, an escape response was reliably triggered, where the worm executed a brief reversal followed by a sharp turn  (Figure~\ref{fig:TurnPartitioning}). The spatial distribution of these escape turns (\textit{i.e.}~sharp turns that immediately follow a reversal) was highly concentrated near the repellent boundary, in stark contrast to that of spontaneous turns upon forward runs (thus without the reversal) that occurred randomly across the arena (Figure~\ref{fig:TurnPartitioning}). 
Therefore, turning behaviour during exploration and escape can be distinguished by the occurrence of reversals and studied independently within the same experiment. Occasionally escape turns were observed far away from the repellent boundary (\textit{i.e.}~$> \SI{7}{\milli\meter}$), but these were infrequent ($14\%$ of all turns $> \SI{7}{\milli\meter}$ from the boundary). As the turn statistics does not allow us to unambiguously associate them with either spontaneous or escape turns, we did not consider them explicitly in our analysis. Similarly, we excluded turns, where the reversal behavior could not be determined or where a reversal occurred immediately after the turn. Finally, we ended up with $5799$ spontaneous turns and with $2943$ escape turns close to the boundary for further analysis. Importantly, this conservative filtering by turn type does not qualitatively affect our conclusions.

Finally, we note that worm behaviour was not significantly perturbed by collisions with other worms within the arena. Collisions did occur at a finite rate, but these were infrequent (on average, once every $\SI{7.8}{\minute}$), could be unambiguously resolved (see Methods) and no effect on crawling speed or sharp turn frequency was detectable beyond the duration of the collision event (Figure~\ref{fig:TurnPartitioning}--Figure Supplement~1). Thus, simultaneous tracking of multiple worms in repellent-boundary arenas enabled efficient acquisition of individual turning statistics across two contrasting behavioral contexts.

\subsection{Variability in spontaneous sharp turn behaviors impact spatial exploration}

The most prominent reorientation behaviours during the worm's exploration of free space are spontaneous sharp turns, which are executed at random times even in the absence of environmental stimuli \cite{Pierce1999,Srivastava2009}. These turns involve a deep body bend, which can occur in either the dorsal or ventral direction, but with a strong statistical bias towards the ventral side \cite{Croll1975,Gray2005}.
Sharp turns can be further classified into $\Omega$- and $\delta$-turns, on the basis of postural statistics \cite{Broekmans2016}. $\Omega$-turns are deep body bends, rendering the worm's shape reminiscent of the greek letter $\Omega$. Compared to $\Omega$-turns, $\delta$-turns exhibit even higher body-bend amplitudes, such that the strongly curved body not only touches but also crosses itself, resulting in larger reorientation angles.

\begin{figure}[t]
\centerline{\includegraphics[width=\textwidth]{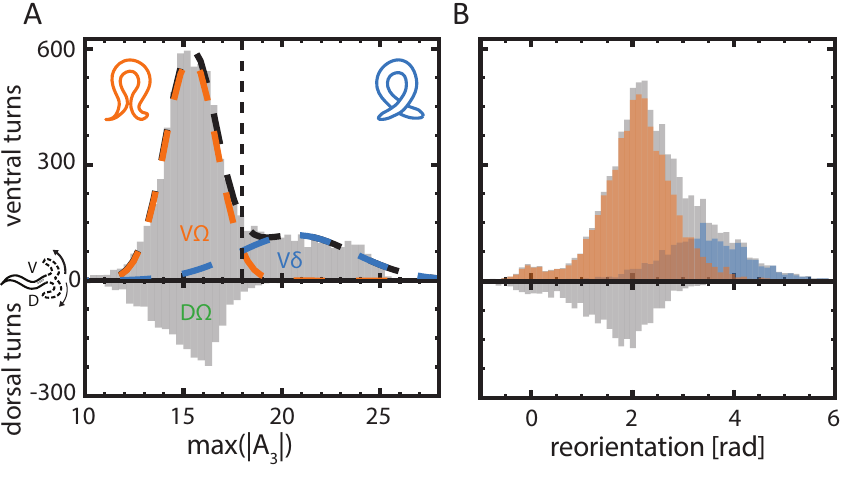}}
\caption{\textbf{$\Omega$ and $\delta$ turns are separated by thresholding the loading of the third Eigenworm} \cite{Broekmans2016,Stephens2008}. (A) The distribution of (top) ventral and (bottom) dorsal maximum Eigenmode loadings across both spontaneous sharp turns and escape turns. The ventral distribution can be approximated as the sum of $2$ Gaussians (black). The orange Gaussian is an approximation of the $\Omega$ turn distribution, and the blue Gaussian is an approximation of the $\delta$ distribution. The $A_3$ value where the lines cross, $18.0$ (black dashed line), is henceforth used as the threshold to separate $\Omega$ and $\delta$ turns. The dorsal distribution only includes $\Omega$ turns. (B) The distribution of (top) ventral and (bottom) dorsal reorientation angles towards the direction of the body bend. The distribution of $\Omega$ and $\delta$ turns after thresholding the maximum $A_3$ amplitude in orange and blue respectively.}
\label{fig:DeltaOmegSeparation}
\end{figure}

To correctly assign sharp turns into these categories, we analysed the postural dynamics throughout each turn. Because $\delta$-turns have been shown to occur only in the ventral direction \cite{Broekmans2016}, we assigned the ventral orientation of each worm to the direction that demonstrated a greater extent in the body-bend amplitude distribution (see Methods). After this D-V orientation assignment, the distribution of body-bend amplitudes for the dorsal direction demonstrated a single peak corresponding to dorsal $\Omega$-turns, whereas that for the ventral direction was better described as a sum of two overlapping peaks, with one corresponding to ventral $\Omega$-turns and the other to ventral $\delta$-turns (Figure~\ref{fig:DeltaOmegSeparation}A). Sorting all ventral turns into bins corresponding to each of these peaks further confirmed that $\delta$-turns do indeed, on average, generate larger angle changes $\Delta\theta$ in the worm's trajectory (Figure~\ref{fig:DeltaOmegSeparation}B).

\begin{figure}[!t]
\centerline{\includegraphics[width=\textwidth]{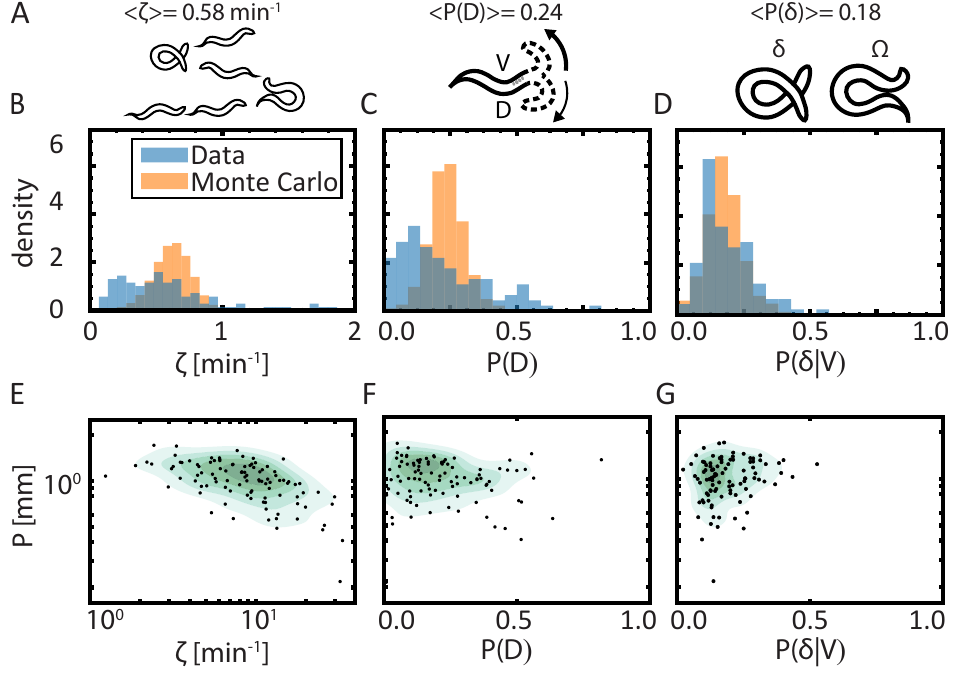}}
\caption{\textbf{Variability in turn frequency, but not D-V bias or $\Omega$-$\delta$ bias, substantially affects exploratory propensity.}
(A) Schematic illustration and population-average statistics of the three random variables that govern spontaneous sharp turns: (left) the turn frequency, $\zeta$; (center) the dorsal turn probability $P(D)$, a measure of D-V bias in sharp turn orientation; (right) the $\delta$-turn probability $P(\delta|V)$, a measure of $\Omega$-$\delta$ bias for ventral turns. The most common turn type is an $\Omega$-turn in the ventral direction.
(B-D) Distribution across the population of the three random variables $\zeta$ (B), $P(D)$ (C), and $P(\delta|V)$ (D) indicate substantial variability across individuals. Blue bars represent statistics for individual measured trajectories, and orange bars are from Monte Carlo simulations assuming all individuals are sampled from a population with identical parameters for the corresponding random variable (see Methods).
(E-G) Relationship between the trajectory persistence length $P$, a measure of exploratory propensity, and the three random variables $\zeta$ (F), $P(D)$ (G), and $P(\delta|V)$ (H). $P$ demonstrates a substantial negative correlation with the sharp turn rate $\zeta$ ($-0.59 \pm 0.12$, $95\%$ CI; $p \le 7.7\cdot10^{-11}$), but its correlation with D-V bias and $\Omega$-$\delta$ bias was, respectively, insignificant ($-0.13 \pm 0.25$, $95\%$ CI; $p \le 2.1\cdot10^{-1}$) and marginally significant ($-0.20 \pm 0.14$, $95\%$ CI, $p \le 4.2\cdot10^{-2}$). Indicated $p$-values were computed using a $t$-test assuming a two-tailed probability.}
\label{fig:figure1}
\figsupp[Interval distribution of spontaneous turns.]{The interval distribution of spontaneous turns for (blue) all worms and (black) individuals.}{
\centerline{\includegraphics[width=\textwidth]{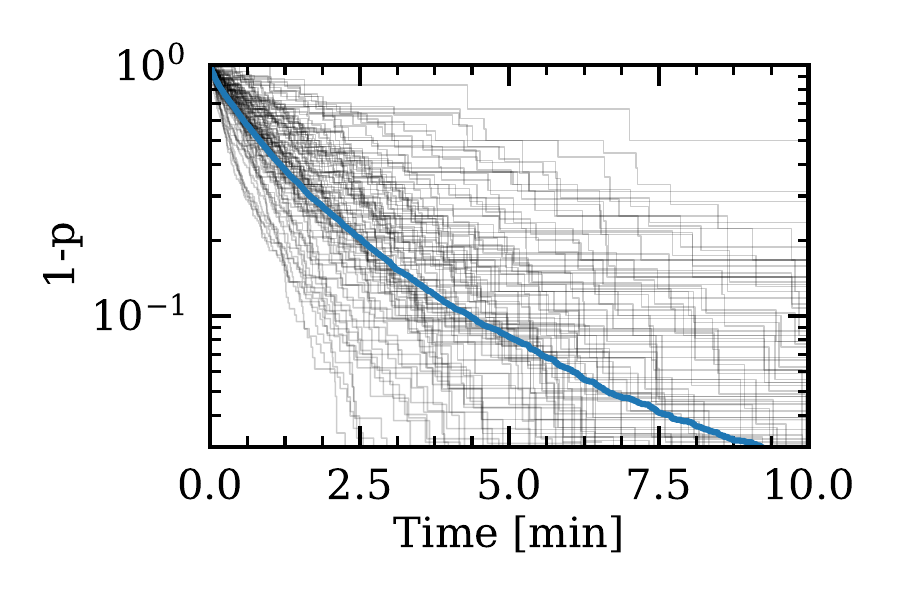}}}
\figsupp[Variability across worms compared to random resampling.]{The observed variability across worms is significantly larger compared to random resampling using population averaged statistics. (A) The standard deviation of (red) the population average turn frequency and (blue) the standard deviation of resampled statistics using the interval distribution (Figure~\ref{fig:figure1}--Figure Supplement~1).
(B) The standard deviation of the measured dorsal turn probability, weighted by the total number of spontaneous turns, is significantly larger compared from random sampling using a coin-flip model using the population average statistics.
(C) The standard deviation of the measured delta turn probability, weighted by the total number of ventral turns, is significantly larger compared from random sampling using a coin-flip model using the population average statistics.}{\centerline{\includegraphics[width=\textwidth]{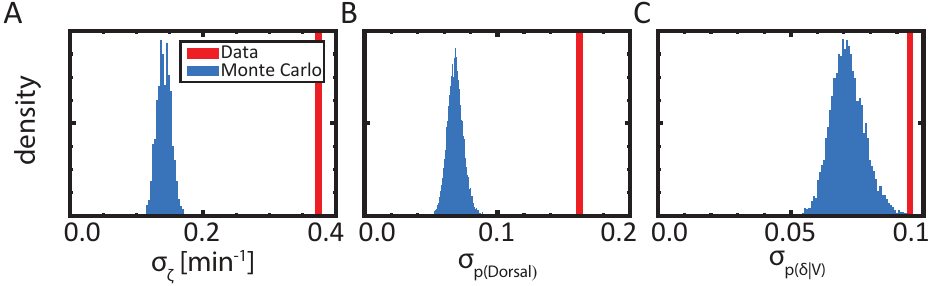}}} 
\figsupp[Variance attributed to individual experience and batch effects.]{The portion of the variance in the measurements that can be attributed to individual experience (blue) and batch effects (orange). The remaining part is attributed to the stochastic nature of the process.
Individual experience is estimated by sampling either the interval distribution in the case of turn frequency or sampling from a binominal distribution in the case of dorsal-$\delta$-turns, where the probabilities are sampled from pooled data from the same batch. This way, batch effects and stochastic effects are included, while individual effects are removed. The relative change in variance is referred to as the individual contribution. The sampling process is repeated to estimate the uncertainty. 
To estimate the effect of batches, first the mean of each batch is subtracted and subsequently total variance is estimated (reducing the degrees of freedom by the number of batches). This is compared against the total variance without subtracting batches. $95\%$ confidence intervals are obtained by bootstrapping for batches. The fraction for variance not accounted for can be attributed to variability as a result of the stochastic processes.}{
\centerline{\includegraphics[width=0.8\textwidth]{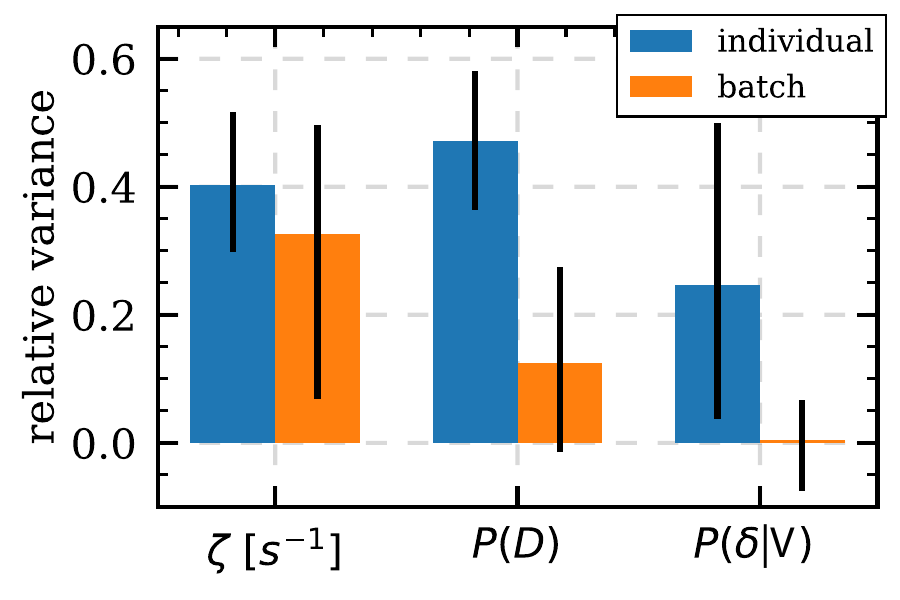}}}
\figsupp[Mean-squared displacement of individual worms and population average.]{The mean-squared displacement of the individual worms (black) and their population average (blue) as function of the trajectory length is ballistic for short length-scales, then becomes diffusive proportional to the persistence length, and finally saturates due to the confinement of the arena.}{
\centerline{\includegraphics[width=0.8\textwidth]{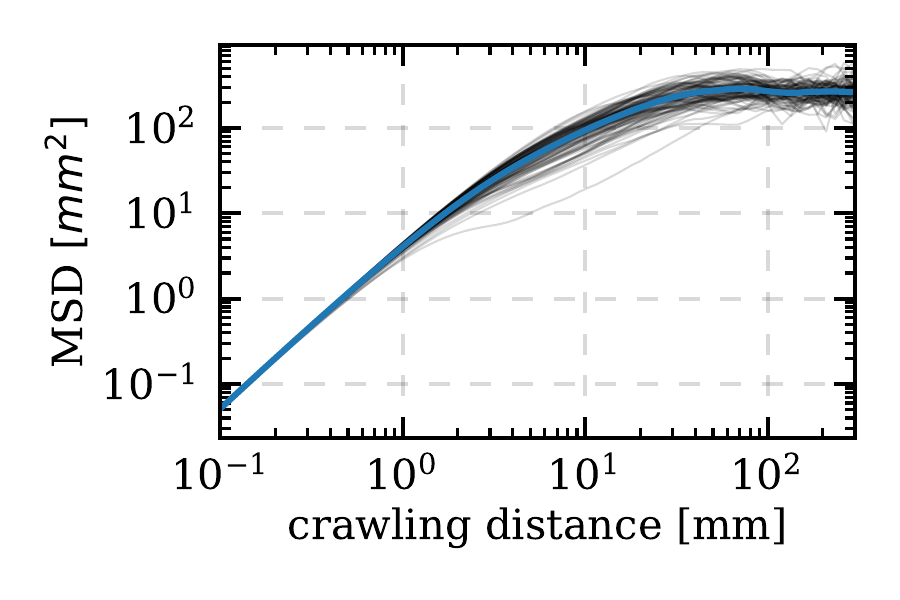}}}
\end{figure}

In addition to the angle change $\Delta\theta$, three random variables suffice to characterize the statistics of sharp turns (Figure~\ref{fig:figure1}A): (1) the rate $\zeta$ ($\SI{}{\minute}^{-1}$) at which sharp turns are executed, (2) the probability $P(\text{D})$ that an executed sharp turn is in the dorsal direction, and (3) the probability $P(\delta|V)$ that an executed ventral sharp turn is a $\delta$-turn. The statistics aggregated over all $100$ worms confirmed that during exploration, there is a strong preference for ventral turns over dorsal turns, and within ventral turns for $\Omega$- over $\delta$-turns (Figures~\ref{fig:DeltaOmegSeparation} and \ref{fig:figure1}) \cite{Gray2005,Kato2015,Nagy2015}.
However at the level of individual trajectories, all three random variables demonstrated substantial variation about the population (Figure~\ref{fig:figure1}B-D, blue). Across individuals, the sharp turn frequency was found to vary by at least an order of magnitude. Some worms did not execute any dorsal turns or $\delta$-turns over the course of the $\SI{2}{\hour}$ measurement, whereas at the opposite extreme, such turns accounted for more than half of all sharp turn events. The variability across individuals was not trivially explained by the finite number of samples within the $\SI{2}{\hour}$ measurement, as the distribution of all three random variables from the measured data were broader than that for Monte Carlo simulations (see Methods) that assumed all worms were realizations of identical random variables with parameters corresponding to the population mean (Figure~\ref{fig:figure1}B-D, orange, Figure~\ref{fig:figure1}--Figure supplement~2).

To evaluate how each of these sharp-turn parameters affect exploration performance, for each trajectory we computed its persistence length $P=D_{\text{t}}/s$, where $D_{\text{t}}$ is the translational diffusion coefficient and $s$ is the movement speed. While $D_{\text{t}}$ depends on both the path geometry and the speed along the trajectory, the persistence length mainly captures the effect of the path geometry and thus more directly links to turning behavior. We extract the persistence length from the relationship between the mean-squared displacement (MSD) and the contour length (Figure~\ref{fig:figure1}--Figure Supplement~4, see Methods).
Comparing the persistence length to the turning parameters for each trajectory revealed that the D-V sharp turn bias was not significantly correlated ($-0.13 \pm 0.25$, $95\%$ CI) to the persistence length, and the $\Omega$-$\delta$ preference was only weakly correlated ($-0.20 \pm 0.14$, $95\%$ CI) (Figure~\ref{fig:figure1}FG).
Thus, neither of these sharp-turn biases substantially affect exploration propensity.
By contrast, the temporal frequency of sharp turns demonstrated a strong negative correlation with the persistence length ($-0.59 \pm 0.12$, $95\%$ CI); worms that turn at higher frequency explore smaller areas (Figure~\ref{fig:figure1}E). In short, we found that each of the random variables ($\zeta$, $P(D)$, and $P(\delta|V)$) vary significantly among individuals, but while sharp turn frequency is strongly negatively correlated with the persistence length, the D-V and $\Omega$-$\delta$ biases have little impact on this measure of exploratory propensity.

\subsection{Bias and fluctuations in gradual turns negatively impact exploration}
\begin{figure}[p!]
\centerline{\includegraphics[width=\textwidth]{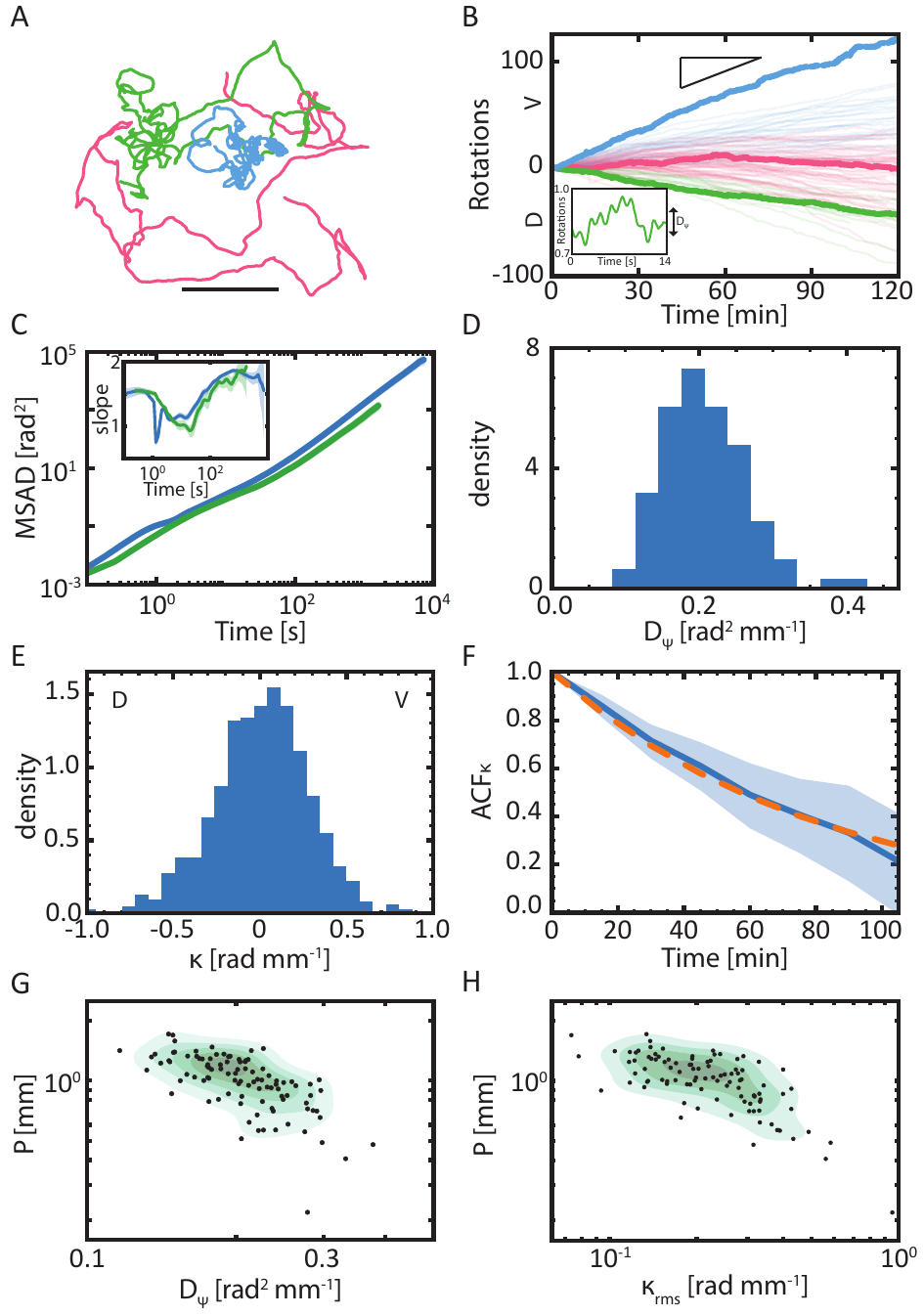}}
\caption{(Caption is on the next page.)}
\label{fig:figure2}
\end{figure}

During exploration, gradual adjustments in the direction of movement occur between spontaneous sharp turn events \cite{Peliti2013,Salvador2014}, causing meandering trajectories on length scales longer than the body wave and in some instances even forming loops (Figure~\ref{fig:figure2}A).
Such effects on trajectories shape reduce the directional persistence of the worm's motion between sharp turn events, and hence could be expected to negatively impact the exploration propensity $P$. We therefore proceeded to quantify the effects of these more subtle reorientations.

\addtocounter{figure}{-1}
\begin{figure} [t!]
\caption{(Previous page.) \textbf{Gradual turning behaviour during exploration demonstrates both short-time fluctuations and long-time biases.} (A) Representative trajectory segments ($10,000$ frames $=\SI{14.5}{\minute}$) for $3$ individual worms, demonstrating strong (blue), intermediate (green) and weak (red) gradual-turn bias and correspondingly different trajectory curvature. In addition to the 'loopiness' caused by the long-timescale bias, diffusive orientation fluctuations induce wiggles in the shape of trajectories. Scale bar $\SI{10}{\mm}$.
(B) Gradual-turn bias can cause trajectories to accumulate many net rotations during the course of the experiment, resulting in a slope. A positive value means that the worm has rotated more in the ventral direction than in the dorsal direction. Inset: angular changes on short time scales from undulatory fluctuations, result in an effective angular diffusion $D_{\psi}$.
(C) The average mean-square angular displacement (MSAD) and (inset) the local exponent (\textit{i.e.}~log-log slope) of the unwrapped average body angle across worms of our data set (blue) and a previously published data set from ref. \cite{Stephens2010} (green; see Methods) show near ballistic behaviour on long time-scales. A slope of $1$ indicates diffusive angular dynamics, and a slope of $2$ corresponds to ballistic angular dynamics. The dip in the slope of the blue curve of the MSAD at $\approx\SI{2}{\second}$ can be attributed to angular oscillations due to the body wave (and is not observed in the green curve, due to differences in the sampling rate and the manner in which angular dynamics were extracted; see Methods).
(D) Probability density histogram of the angular diffusion coefficient $D_\psi$, extracted from each of 100 individual trajectories.
(E) Probability density histogram of the local gradual-turn bias $\kappa$, defined as the average trajectory curvature within $\SI{15}{\minute}$ windows, extracted from all such non-overlapping windows in all 100 trajectories. The sign of $\kappa$ was set to be positive in the ventral (V) direction and negative in the dorsal (D) direction. The average rotational drift for each worm shows no dorso-ventral population mean.
(F) Slow fluctuations gradual-turn bias decorrelate on a timescale comparable to the duration of the measurement, and can be fit by a single-exponential decay with a time constant of $82 \pm \SI{17}{\minute}$. Because the time scale of the fluctuations is similar to the length of the measurement, the mean cannot be established of individual measurements, and the global mean value of $0$ is used.
(G) The angular diffusivity $D_\psi$ is negatively correlated with the persistence length $P$, with a correlation coefficient $-0.57 \pm 0.13$ ($95\%$ CI) ($p<2.9\cdot 10^{-10}$).
(H) The root-mean-square gradual-turn bias $\kappa_\text{RMS}$ is strongly negatively correlated with the persistence length $P$, with a correlation of $-0.72 \pm 0.11$, ($95\%$ CI) ($p<1.3\cdot 10^{-17}$). Indicated $p$-values were computed using a two-tailed $t$-test.}
\figsupp[Reorientations are nearly decorrelated after a single body wave.]{Reorientations are nearly decorrelated after a single body wave. To eliminate the effect of the body wave oscillations, the orientation $\psi$ was evaluated every body wave at the same body posture, computed from the phase of the first $2$ Eigenworms \cite{Stephens2008}. $\Delta\psi$ is the difference of $\psi$ after exactly $1$ body wave. 
The distribution flattens at a value slightly greater than $1$, due to the rotational bias.
Interval distributions show the $95\%$ confidence interval of the mean across all worms.}{
\centerline{\includegraphics[width=0.8\textwidth]{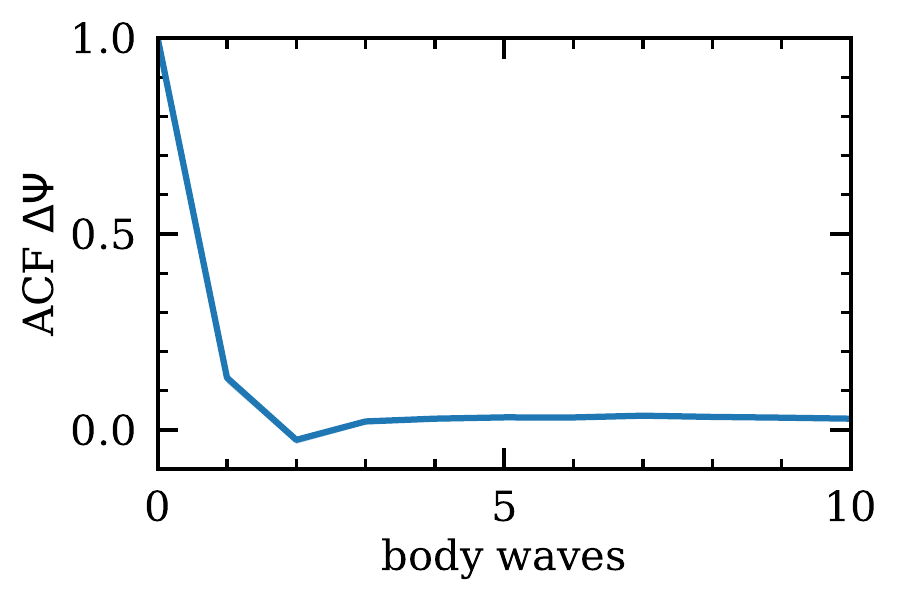}}}
\end{figure}

Gradual reorientation dynamics were extracted from the worm's body orientation (see Methods), which provides an accurate proxy for the direction of movement (\textit{i.e.} the velocity bearing) during runs, even when moving at low speeds. Over the course of the experiment, the unwrapped body orientation (\textit{i.e.} cumulative angle change) of most worms demonstrated many full rotations, indicating a significant bias in the gradual-turn dynamics (Figure~\ref{fig:figure2}B). Interestingly, whereas some worms accumulated rotations persistently in the dorsal (14/100 worms) or ventral (24/100) direction (Figure~\ref{fig:figure2}B, green and blue curves, respectively), most worms (62/100) demonstrated rotations in both the dorsal and ventral directions (Figure~\ref{fig:figure2}B, red curves).

The gradual-turn dynamics also feature faster orientation fluctuations, causing the unwrapped body-orientation trajectories to be jagged, rather than smooth curves (Figure~\ref{fig:figure2}B). To examine the statistics of gradual turns across timescales, we computed the mean-squared angular displacement (MSAD) of the unwrapped body orientation as a function of time (Figure~\ref{fig:figure2}C, blue curve), which showed similar dynamics compared to a previously published data set \cite{Stephens2010} (Figure~\ref{fig:figure2}C, green curve).
For timescales longer than the body wave oscillations ($\sim \SI{2}{\second}$) but less than $\sim 10 \si{\second}$, the local power-law scaling exponent (\textit{i.e.} the log-log slope) of $\text{MSAD}(t)$ was close to unity (Figure~\ref{fig:figure2}C, Inset), indicating that fluctuations around this timescale are well-approximated as pure (\textit{i.e.} Brownian) angular diffusion. Consistently, angular deviations observable at short times/lengths decorrelated nearly completely within a single body wave (Figure~\ref{fig:figure2}--Figure Supplement~1). Therefore, these random orientation changes on short time- and length-scales (comparable to the body wave) can be modeled as an angular diffusion process \cite{Helms2019} (see Methods), characterized by a angular diffusion coefficient $D_{\psi}$ which varied from worm to worm but most commonly was around $\SI{0.2}{\radian^2\milli\meter}^{-1}$ (Figure~\ref{fig:figure2}D). 
For longer timescales, the angular dynamics were increasingly super-diffusive, with the local scaling exponent plateauing around $\SI{1000}{\second}$ at a maximum value of $\sim 1.86$ (Figure~\ref{fig:figure2}C, Inset), consistent with a persistent gradual-turn bias (perfectly circular trajectories would yield an exponent of $2.0$).
Thus, the worms' reorientation dynamics can be described as a combination of two processes: (1) angular diffusion resulting in meandering trajectories and (2) a gradual-turn bias (rotational drift) causing the trajectories to form loops \cite{Helms2019}.

To quantify the gradual-turn bias, we computed the trajectory curvature $\kappa$, averaged over windows of $\SI{15}{\minute}$, the time scale around which the slope of $\text{MSAD}(t)$ was maximal (Figure~\ref{fig:figure2}C, Inset) and hence orientation dynamics were most persistent. The distribution of $\kappa$ pooled from all 100 trajectories was distributed approximately symmetrically in the dorsal and ventral directions (Figure~\ref{fig:figure2}E) with a mean that is not significantly different from zero ($0.00 \pm 0.02$, $95\%$ CI), indicating no net D-V bias across the ensemble of trajectories. However, as noted above, the gradual-turn bias of individual trajectories tends to fluctuate slowly over time. To characterize the timescale of such slow fluctuations in the gradual-turn bias, we computed the autocorrelation function of $\kappa$ ($\text{ACF}_\kappa(t) \equiv \langle \kappa(0)\kappa(t)\rangle$) (Figure \ref{fig:figure2}F).
Within the timescale of the measurement, the ensemble average $\langle\text{ACF}_\kappa(t)\rangle$ decayed to $0.2$, with a profile consistent with a single-exponential $\langle\text{ACF}_\kappa(t)\rangle\sim e^{-t/\tau}$ with decay time constant $\tau = 82 \pm 17$~$\si{\minute}$ (see methods). Thus, although the $2 \si{\hour}$ timescale of our experiment precludes confirming the full decorrelation of these slow fluctuations, the data are compatible with a model in which gradual-turn bias fluctuates slowly in a manner that when averaged over very long times ($\gg \tau$) has little or no net bias (per the near-zero $\kappa$-distribution mean), but has at any instant of time a finite magnitude (in the range of the $\kappa$-distribution width).

As noted above, both the short-timescale angular diffusion and the longer-timescale bias of gradual turns can be expected to negatively impact exploration. Consistent with this idea, worm-to-worm variation in both the angular diffusivity $D_\psi$ and the root-mean-square (RMS) magnitude of the gradual-turn bias ($\kappa_{\text{RMS}}$) demonstrated clear negative correlations with the exploration propensity $P$ of trajectories (Figure ~\ref{fig:figure2}G,H).

In summary, although the gradual-turn bias is relatively weak ($\langle\kappa_{\text{RMS}}\rangle\approx 0.2\si{\radian\milli\meter}^{-1}$) and fluctuates slowly over time with an approximate zero mean, its cumulative effect on the exploration propensity $P$ over long time- and length-scales are comparable to, or greater than, that of the short time- and length-scale angular diffusion.

\subsection{A finite gradual-turn bias leads to an optimal choice for angular diffusivity}
The data presented above demonstrate that each of three turning-behavior parameters --- the sharp turn rate $\zeta$, gradual-turn diffusivity $D_\psi$ and bias magnitude $\kappa_\text{rms}$ are negatively correlated with the exploratory propensity $P$ (\textit{i.e.} the persistence length) of trajectories. Yet the manner in which these different turning components affect exploration might not be independent. To gain insight into the relative contributions of, and interactions between, these parameters in determining the persistence length $P$, we constructed a minimal model of the worm's turning behaviour.

Changes in the lab-frame orientation $\psi$ along the trajectory contour $x$ is described as a sum of two terms: a constant angular drift accounting for the gradual-turn bias, and angular diffusion:
\begin{equation} d \psi (x) = \kappa dx + \sqrt{2D_{\psi}}dW_x. \label{eq:angulardiffusion} \end{equation}
where $\kappa$ [$\SI{}{rad\per\milli\meter}$] is the constant angular drift, $D_{\psi}$ [$\SI{}{rad^2\per\milli\meter}$] is the angular diffusion coefficient, and $dW_x$ is an increment of a Wiener process along the contour coordinate $x$.
Sharp reorientation events occur with a uniform probability per unit length along the trajectory contour, determined by the turn frequency $\zeta \ [\SI{}{\per\second}$] and speed $s$ [$\SI{}{\milli\meter\per\second}$] of the worm:
\begin{equation}p_{\text{turn}}=\zeta / s \,dx.\label{eq:TurnRate}\end{equation}
The experimental observation that sharp turns do not completely randomize the orientation but are instead biased, on average, slightly towards the reverse direction (\textit{i.e.} $\langle\cos\Delta \theta\rangle<0$; Figure~\ref{fig:DeltaOmegSeparation}B) can be accounted for by scaling the sharp turn frequency by a factor $\alpha=1- \langle \cos \Delta \theta \rangle$ \cite{Taktikos2013,Locsei2007}, for which we use the population averaged value $\alpha=1.33$.

\begin{figure}[t!]
\centerline{\includegraphics[width=\textwidth]{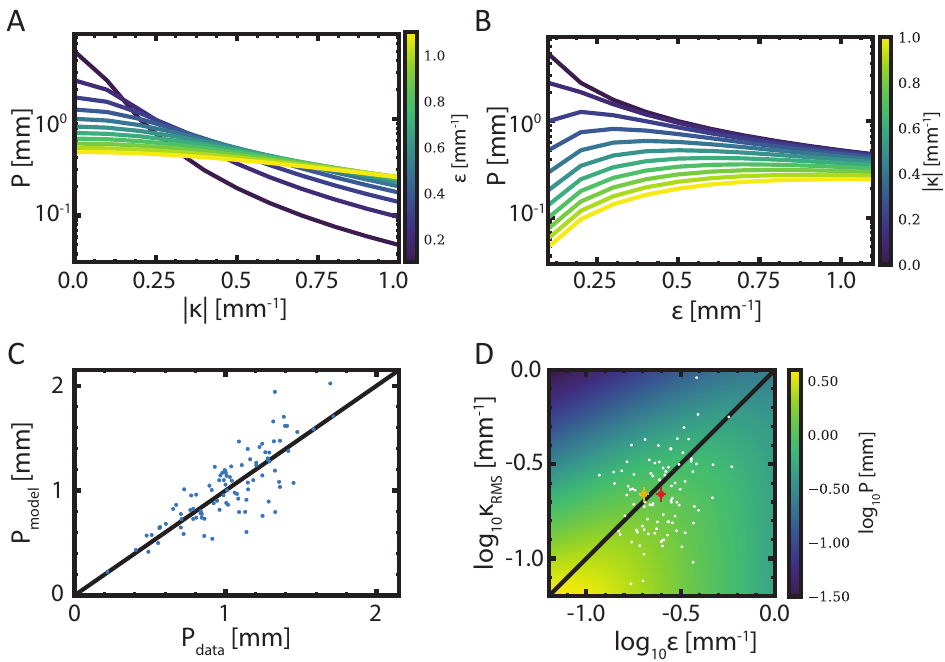}}
\caption{\textbf{A simple turning model explains the data and reveals an optimality principle for exploration under biasing constraints}
(A) Exploratory propensity, characterized by the trajectory persistence length $P$, decreases monotonically with increasing turning bias $\kappa$, regardless of the rate of random reorientations $\epsilon$.
(B) By contrast, $P$ can either increase or decrease with $\epsilon$, depending on the value of $\kappa$.
(C) The measured persistence length $P_\text{data}$ agrees well with predictions of the model $P_\text{model}$ based on the turning parameters $\kappa_\text{RMS}$ and $\epsilon$ measured in each worm. The analytical model assumes a constant $|\kappa|=\kappa_\text{RMS}$.
(D) The magnitude of the gradual-turn bias $\kappa_\text{RMS}$ and the effective random-reorientation rate $\epsilon$ is of the same order. Each trajectory is displayed as a white point. The red point is the population average, computed from all trajectories. The orange point indicates the population average for the case that sharp turns are ignored (equivalent to the limit $\alpha \rightarrow 0$ in our model), so that $\epsilon$ is defined by rotational diffusion alone (\textit{i.e.} $\epsilon \rightarrow D_\psi$). Error bars represent $95\%$ confidence intervals. The analytical expression for the persistence length is 
$ P= \epsilon/(\kappa^2+\epsilon^2)/2$, where $\kappa$ is a constant angular drift (thus $|\kappa|=\kappa_\text{RMS}$) and 
$\epsilon=\alpha\zeta/s+D_{\psi}$ is the effective reorientation rate with the angular diffusion coefficient $D_{\psi}$, the speed $s$, the sharp turn frequency $\zeta$, and a factor $\alpha$ accounting for the non-uniform distribution of sharp turn angles.
}
\label{fig:figure3}
\figsupp[The analytic solution of the model closely follows simulations.]{The analytic solution of the model closely follows simulations. Simulations are performed with constant speed ($s=\SI{0.15}{\milli\meter\per\second}$) and $2 \cdot 10^6$ data points at $\SI{2}{\hertz}$, using the orientational dynamics described in equation (\ref{eq:angulardiffusion}) and (\ref{eq:TurnRate}). A large space of motility parameters has been simulated that includes that of the measurements. $\kappa$ and $\zeta/s$ have been varied from $\SI{0}{\per\milli\meter}$ to $\SI{1}{\per\milli\meter}$ in steps of $\SI{0.2}{\per\milli\meter}$. $D_{\psi}$ has been varied from $\SI{0.1}{\per\milli\meter}$ to $\SI{0.9}{\per\milli\meter}$ in steps of $\SI{0.2}{\per\milli\meter}$. Sharp turn are modelled as a complete randomization of the reorientation. The persistence length extracted from the simulated trajectories is practically identical (correlation of $99.8\%$) to that computed from the model (equation (\ref{eq:persLength})).}{
\centerline{\includegraphics[width=0.8\textwidth]{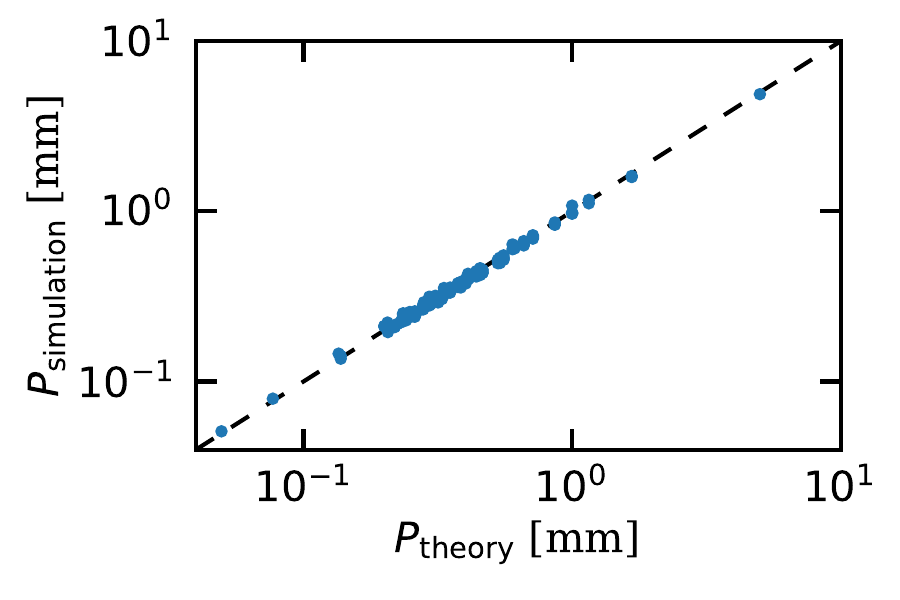}}}
\figsupp[Evidence for the existence of small reorientation events.]{The reorientation distribution is well fitted by $2$ Gaussians, which may indicate the existence of small reorientations. The orientation has been evaluated after subsequent body waves during runs at similar body postures (evaluated from the first $2$ Eigenworms \cite{Stephens2008}), to ignore the effect of the oscillatory motion. (A) The reorientation angle $\Delta\psi$ is well fitted as the sum of $2$ Gaussians (orange curve), with a standard deviation of $\SI{13.5}{\degree} \pm \SI{0.1}{\degree}$ and $\SI{28.3}{\degree} \pm \SI{0.4}{\degree}$ with mean values of $\SI{4.3}{\degree}\pm \SI{0.9}{\degree}$ and $\SI{-2.8}{\degree} \pm \SI{0.3}{\degree}$, respectively (yellow curves). Fits are performed with the lmfit package in python using the Levenberg-Marquardt method. (B) Four exemplary (left) time series and (right) centroid trajectories of potential shallow turns with a reorientation angle $>\SI{45}{\degree}$. To compute the curvature rate, the average body angle of the worm is evaluated at equally spaced distances of $\SI{20}{\micro\meter}$ and the derivative is computed using a Savitzky–Golay filter ($3$rd order with a window size $\SI{300}{\micro\meter}$).}{
\centerline{\includegraphics[width=\textwidth]{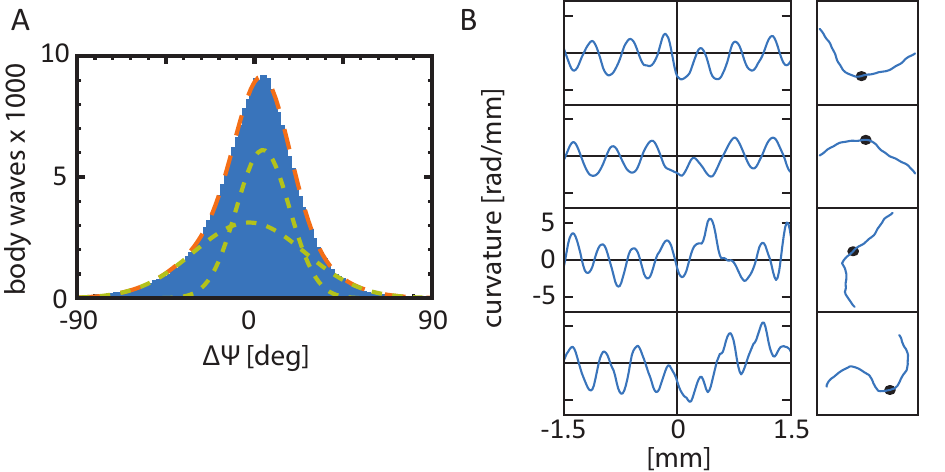}}}
\end{figure}

We can solve analytically for the expected persistence length $P$, yielding a simple closed form expression (see SI text):
\begin{equation}P= \frac{1}{2}\frac{\epsilon}{\kappa^2+\epsilon^2}.
\label{eq:persLength}\end{equation}
Here $\epsilon=\alpha\zeta/s+D_{\psi}$ is the effective rate of random reorientation, combining the effects of sharp turns and angular diffusion. The persistence length computed using this analytical solution is in excellent agreement with numerical simulations of the model (Figure~\ref{fig:figure3}--Figure Supplement~1), and further accurately predicts the persistence length of the measured worm trajectories (Figure \ref{fig:figure3}C; Pearson correlation $\approx 0.8$).

The simple analytical form of Eq.~(\ref{eq:persLength}) also provides insights into the dependence of this measure of exploratory propensity on the underlying parameters. It is clear that a non-zero curvature ($|\kappa|>0$) monotonically impairs the persistence length (\textit{i.e.} $\partial P/\partial|\kappa|<0$; Figure \ref{fig:figure3}A,B). Interestingly, it also reveals that, given any nonzero value of $\kappa$, there exists a finite value of $\epsilon$ that maximizes $P$ (\textit{i.e.} $\partial^2 P/\partial\epsilon^2<0$ at $\partial P/\partial\epsilon=0$). When $\epsilon \gg |\kappa|$, effects of the gradual-turn bias are negligible and the persistence length becomes inversely proportional to $\epsilon$. When $\epsilon \ll |\kappa|$, the trajectory becomes circular, leading to oversampling of space within a region of length scale $\kappa^{-1}$, the circling radius. In this latter regime, $P$ increases with $\epsilon$ because random reorientations are required to free the worm from the circular orbit. Thus, $P$ increases with $\epsilon$ at small $\epsilon$ but decreases with $\epsilon$ at large $\epsilon$, and an optimum in $P$ occurs when the random reorientation rate is balanced with the curvature (\textit{i.e.} when $\epsilon = \kappa$).

Figure \ref{fig:figure3}D compares the measured values for $\epsilon$ and $\kappa_\text{RMS}$ with the theoretical predictions on optimal exploration. 
Interestingly, although both $\epsilon$ and $\kappa_\text{RMS}$ were found to vary substantially across individuals, when averaged over the entire measured population (Figure \ref{fig:figure3}D, white points), these values were nearly identical to one another (red point) and thus close to the predicted optimum (solid line). This can be largely attributed to the contribution of the rotational diffusivity $D_\psi$, which for exploring worms in the absence of food, evidently dominates over the sharp turn rate $\zeta$ and is by itself comparable in magnitude to $\epsilon$ (Figure \ref{fig:figure3}D, compare orange and red points). Our model therefore indicates that, on average, the rotational diffusivity $D_\psi$ of worms is set very close to the optimal value that balances their finite gradual turn bias $\kappa$ to maximise the exploratory propensity, quantified by $P$.

\subsection{Context-dependent modulation of both D-V and \textOmega{}-\textdelta{} statistics symmetrizes escape performance under biasing constraints}

\begin{figure}[htp]
\centerline{\includegraphics[width=1\textwidth]{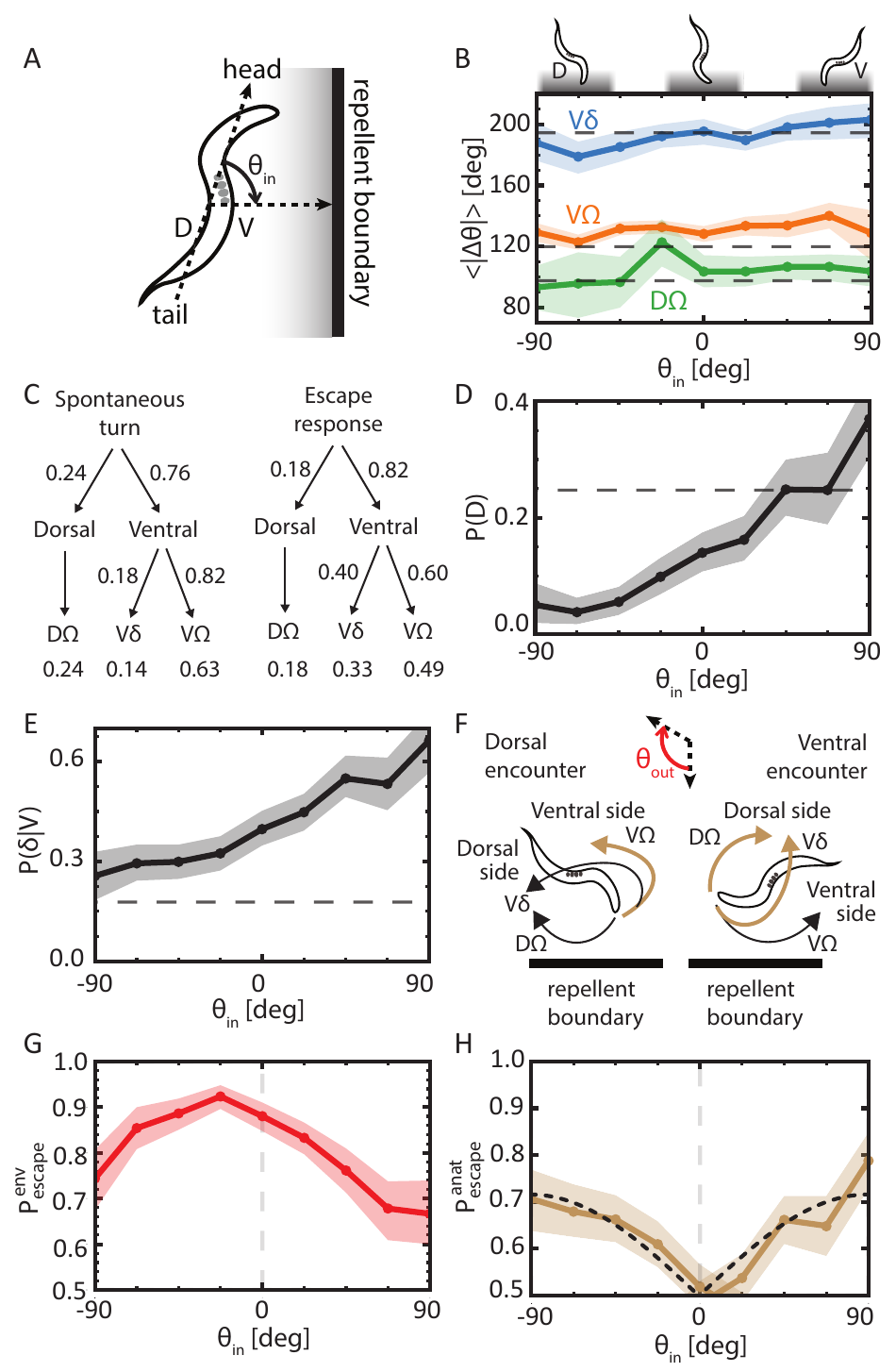}}
\caption{(Caption is on the next page.)}
\label{fig:figure4}
\end{figure}
\addtocounter{figure}{-1}
\begin{figure} [t!]
\caption{(Previous page.) \textbf{Escape-turn statistics reveal discrete, rather than continuous, control of exit angles to overcome biasing constraints.} (A) We characterise worm orientation during escape turns by the angle $\theta_\text{in}$ of the body orientation vector (pointing from tail to head) relative to the repellent gradient (approximated as the vector pointing from the worm centroid to the nearest point on the repellent boundary), where $\theta_\text{in}=\SI{0}{\degree}$ means the worm points directly up the gradient and $\SI{0}{\degree}<\theta_\text{in}<\SI{180}{\degree}$ and $\SI{-180}{\degree}<\theta_\text{in}<\SI{0}{\degree}$ correspond to the nearest repellent boundary being on the ventral and dorsal sides, respectively.
(B) The average reorientation angle $\langle |\Delta\theta |\rangle$ of escape turns close to the boundary demonstrates negligible dependence on $\theta_\text{in}$ for dorsal $\Omega$- (green), ventral $\Omega$- (orange) and ventral $\delta$-turns (blue), respectively. The dashed line denotes the values for spontaneous turns.
(C) Like for spontaneous turns, the worm makes a decision between dorsal and ventral turns, and if turning ventrally between omega and delta turns. The decision tree shows the average probabilities for spontaneous turns ($P(D)=0.24\pm0.01$, $P(\delta|V)=0.18\pm0.1$) and escape responses close to the boundary ($P(D)=0.16\pm0.01$, $P(\delta|V)=0.40\pm0.02$). Escape-like turns far away from the boundary display intermediate values for $P(D)=0.21\pm0.03$ and $P(\delta|V)=0.30\pm0.03$.
(D) The D-V bias of escape turns is modulated such that the dorsal turn probability $P(D)$ is suppressed when the repellent is encountered on the dorsal side ($\SI{-90}{\degree}<\theta_\text{in}<\SI{0}{\degree}$). Dashed line: average $P(D)$ for spontaneous turns during exploration.
(E) The $\Omega$-$\delta$ bias of ventral escape turns is also modulated, with the $\delta$-turn probability $P(\delta|V)$ being increased when the repellent is encountered on the ventral side ($\SI{0}{\degree}<\theta_\text{in}<\SI{90}{\degree}$). Dashed line: average $P(\delta|V)$ for spontaneous turns during exploration.
(F) After the turn, the worm leaves at an angle $\theta_\text{out}$ (red arrow), which is defined in the environmental reference frame, analogously to $\theta_\text{in}$ with $\theta_\text{out}=0$ if the worm is heading directly towards the boundary. In the anatomical reference frame of the moving worm, the escape turn results in an exit angle either on the ventral side (via a ventral $\Omega$-turn) or on the dorsal side (via a dorsal $\Omega$- or a ventral $\delta$-turns) of its body. Successful escape in the anatomical reference frame is defined by turns that result in exiting the turn on the opposite side of the body as the repellent encounter (gold arrows).
(G) The environmental escape probability $P_\text{escape}^{env}$ quantifies how likely the worm is moving moving away from the repellent boundary after the turn (\textit{i.e.}~$|\theta_\text{out}|>\SI{90}{\degree}$), which depends on the incoming angle $\theta_\text{in}$. The distribution is asymmetric and higher if the cue is encountered on the dorsal side of the body.
(H) The escape probability in the anatomical reference frame quantifies whether the worm escapes on the dorsal or ventral side of its incoming body, respectively: $P_\text{escape}^{anat}=(1-P(D))(1-P(\delta|V))$ if $\theta_\text{in}\leq\SI{0}{\degree}$ and $P_\text{escape}^{anat}=P(D)+(1-P(D))P(\delta|V)$ if $\theta_\text{in}\geq\SI{0}{\degree}$. It is modulated depending on the incoming orientation $\theta_\text{in}$. The black dotted curve denotes a fit of $\pm A\sin(\theta_\text{in})+0.5$.
In all panels, shaded regions correspond to $95 \%$ confidence intervals.}
\figsupp[Escape turns are triggered at the boundary.]{Escape turns are triggered when the worm approaches the boundary and reorient the worm away from the boundary. The average distance from the boundary is averaged across escape turns.}{
\centerline{\includegraphics[width=\textwidth]{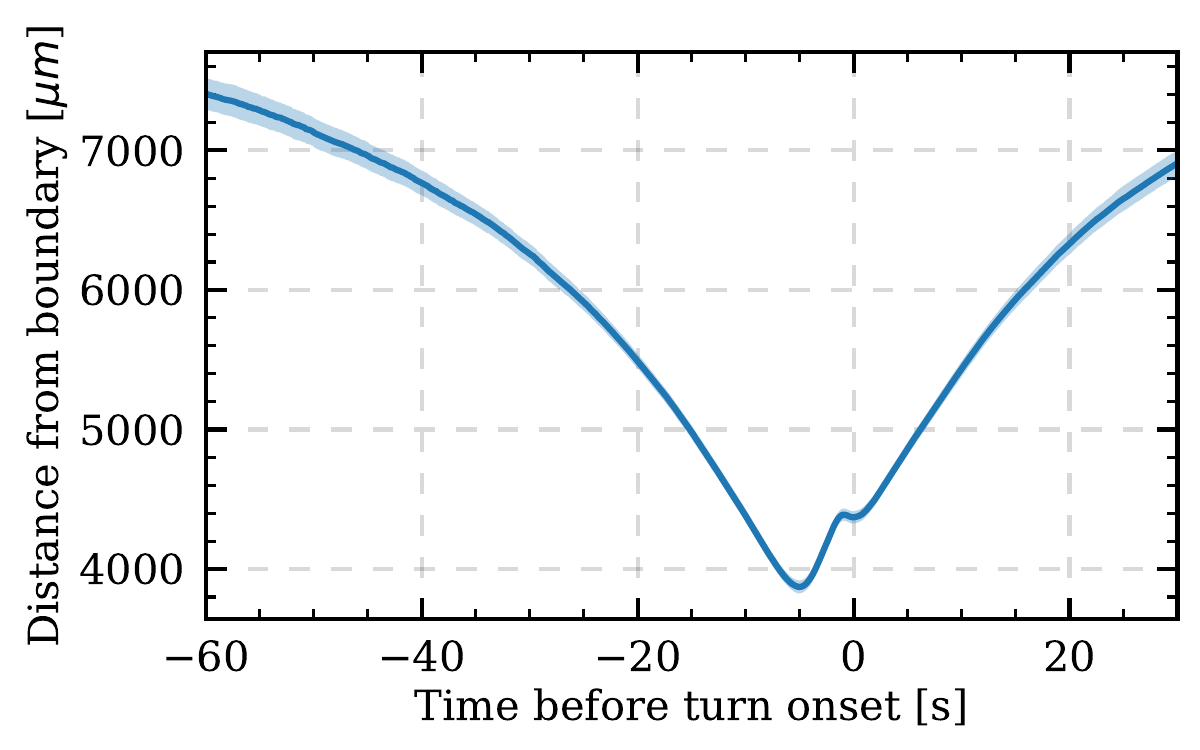}}}
\figsupp[Comparison of escape response and spontaneous turn across individuals.]{Worm variability in (A) dorsal and (B) $\delta$ preference during the escape response and spontaneous turns is not significantly correlated.}{
\centerline{\includegraphics[width=\textwidth]{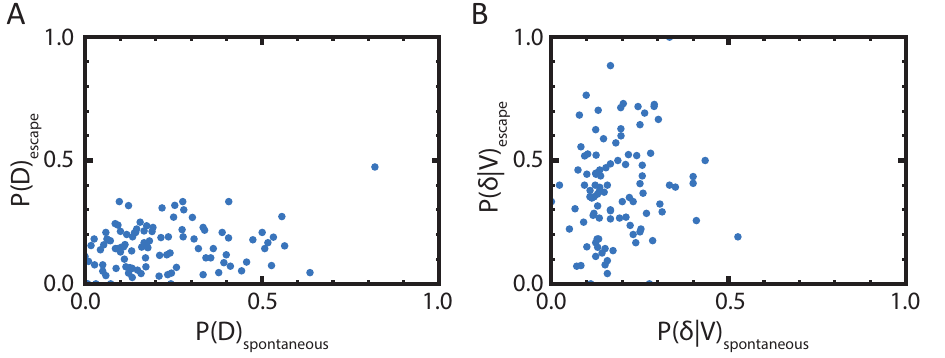}}}
\end{figure}

Escaping from a threat represents an acute challenge for trajectory reorientation. The goal of an escape response is to rapidly change the direction of motion so as to crawl away from the perceived source of an aversive stimulus (within our experiment, the arena boundaries impregnated with a chemorepellent) (Figure~\ref{fig:figure4}--Figure supplement~1). If the worm were able to respond perfectly to the threat, the orientation $\theta_\text{out}$ of the escaping worm after the turn would point in the direction exactly down the repellent gradient, orthogonal to and away from the boundary ($\theta_\text{out}=\SI{180}{\degree}$; we define $\SI{0}{\degree}$ to be the direction of the shortest path to the boundary). However in reality, the re-orientation behavior can be expected to deviate from this optimal case due to various constraining factors, ranging from asymmetries in the worm's anatomy and control physiology to limited information about the orientation of the aversive environmental gradient.

We resolved a total of $2.943$ escape responses within our dataset, triggered near the repellent boundary ($<\SI{7}{mm}$ from the boundary, Figure~\ref{fig:TurnPartitioning}), which consist of a reversal motion followed by one of three types of posturally distinct sharp turns: ventral omega ($V\Omega$), ventral delta ($V\delta$), or dorsal omega ($D\Omega$). This set constitutes the full repertoire of sharp turns also observed in the freely moving case, Figure~\ref{fig:DeltaOmegSeparation}. The amplitude of each of these sharp-turn types, characterized by the typical angle change $\langle|\Delta\theta|\rangle$, were --- if at all --- very weakly modulated as a function of the incoming angle of approach toward the repellent boundary $\theta_\text{in}$ (Figure~\ref{fig:figure4}A,B), and remained very similar to those during free exploration (Figure~\ref{fig:figure4}B, dashed lines).

By contrast, the probability of executing each of the three sharp-turn types differed significantly from that during free exploration. During escape responses, the average dorsal-turn probability $P(D)$ is decreased from $0.24\pm0.01$ to $0.16\pm0.01$, and the average delta-turn probability $P(\delta|V)$ is increased during escape turns from $0.18\pm0.01$ to $0.40\pm0.02$ ($95\%$ CI, using bootstrapping), leading to an overall increase in $V\Omega$ and $V\delta$ turns, and a decrease in $D\Omega$ turns. (Figure~\ref{fig:figure4}C). 
Furthermore, the average dorsal and $\delta$-turn biases of the escape turn were not correlated with the aforementioned individual biases of each worm during exploration (Figure~\ref{fig:figure4}--Figure Supplement~2), thus ruling out that these differences in decision statistics are due to sampling bias (which could arise if individual turn biases were correlated with boundary encounter rates). Taken together, these observations strongly suggest that the worm's decision to select one of the three sharp-turn types during escape reflect the specific context of the escape challenge.

Consistent with this idea, we found that both the probability for a dorsal turn and that for a $\delta$ turn during escape are modulated as a function of the incoming angle $\theta_\text{in}$ of approach to the repellent boundary. When the worm approaches the repellent boundary on its dorsal side ($\theta_\text{in}<\SI{0}{\degree}$), the dorsal turn probability $P(D)$ is strongly suppressed (Figure \ref{fig:figure4}D) below that for spontaneous turns during exploration (dashed line), down to nearly zero as $\theta_\text{in}$ approaches $\SI{-90}{\degree}$. The extent of $P(D)$ modulation is more modest when the repellent is encountered on the ventral side ($\theta_\text{in}>\SI{0}{\degree}$); worms make ventral turns more than half of the time even when $\theta_\text{in}$ approaches $\SI{+90}{\degree}$ (Figure \ref{fig:figure4}D).
The less extensive modulation of the D-V bias for $\theta_\text{in}>\SI{0}{\degree}$ is apparently suboptimal, but we found that it is compensated --- at least in part --- by adjusting the $\Omega$-$\delta$ bias of ventral turns, with $P(\delta|V)$ increasing more than 2-fold across the full range of possible incident angles ($\SI{-90}{\degree}<\theta_\text{in}<\SI{90}{\degree}$) (Figure \ref{fig:figure4}E). Thus, control of escape-turn behavior evidently is achieved through digital, rather than analog logic: what is modulated is not the turn amplitude $|\Delta\theta|$, but rather the probability of selecting from a discrete repertoire of sharp turn types, each of which has an essentially invariant characteristic amplitude.  

But to what end might the worm be executing these discrete decisions? To answer this question, we considered two contrasting performance metrics for the escape behavior (Figure \ref{fig:figure4}F). The first assesses the probability $P_\text{escape}^\text{env}$ of a successful escape in the reference frame of the environment, where success is defined by the exit angle pointing in the downward direction of the repellent gradient. Specifically, we extract the exit angle with respect to the repellent boundary, $\theta_\text{out}$ (Figure \ref{fig:figure4}F, red arrow), and define $P_\text{escape}^\text{env}$ as the probability that the exit angle is pointing away from the repellent boundary, \textit{i.e.} $P_\text{escape}^\text{env}=\text{Prob}\{|\theta_\text{out}|>\SI{90}{\degree}\}$. 
The second metric assesses the probability $P_\text{escape}^\text{anat}$ of successful escape in the anatomical reference frame of the moving worm, where success is defined by the worm exiting the escape turn on the side of its body opposite to that of the repellent-boundary encounter (Figure \ref{fig:figure4}F, gold arrows). That is, $P_\text{escape}^\text{anat}$ is defined as the probability of exiting on the dorsal side upon ventral encounter, and exiting on the ventral side upon dorsal encounter. Both $P_\text{escape}^\text{env}$ and $P_\text{escape}^\text{anat}$ thus reduce escape performance to a single probability defining a binary random variable --- whether the worm exits in the favorable direction or not --- but the reference frame in which "favorable direction" is defined differs between the two.  

The environmental-frame escape probability $P_\text{escape}^\text{env}$ was found to be significantly favorable compared to the 50-50 odds expected if control is absent ($P_\text{escape}^\text{env}$>0.5), across the full range of encounter angles $\theta_\text{in}$ (Figure \ref{fig:figure4}G). However, $P_\text{escape}^\text{env}$ does decrease for large absolute values of $\theta_\text{in}$. While for $\theta_\text{in}\approx\SI{0}{\degree}$, essentially any choice among the three turn types ($V\Omega$, $V\delta$, and $D\Omega$) leads to an escape away from the repellent boundary (because their typical amplitudes $\langle|\Delta\theta|\rangle$ are all greater than $\SI{90}{\degree}$; Figure \ref{fig:figure4}B), for $|\theta_\text{in}|\approx\SI{90}{\degree}$ the choice of turn type is important for successful escape in the environmental frame (\textit{i.e.}~to achieve $|\theta_\text{out}|>\SI{90}{\degree}$). The reduced performance near $|\theta_\text{in}|\approx\SI{90}{\degree}$ thus indicates a significant fraction of escape trials result in the "wrong" choice among the sharp-turn repertoire --- resulting in failed escape --- perhaps due to the limited precision at which sensory measurements of the chemical gradient can be made within the short duration ($\sim\SI{10}{\second}$) of the escape response. And perhaps reflecting the underlying bias in sharp-turn selection probabilities (Figure \ref{fig:figure4}B), environment-frame performance as a function of the boundary-encounter angle $\theta_\text{in}$ is asymmetric about $\theta_\text{in}=\SI{0}{\degree}$ (Figure~\ref{fig:figure4}G), with a higher escape performance when worms encounter the repellent boundary on the dorsal side of their anatomy. 

By contrast, the anatomical-frame escape probability $P_\text{escape}^\text{anat}$ was found to be approximately symmetric about $\theta_\text{in}=\SI{0}{\degree}$, increasing from $P_\text{escape}^\text{anat}\approx0.5$ when encountering the boundary head on ($\theta_\text{in}\approx\SI{0}{\degree}$), and increasing to $P_\text{escape}^\text{anat}\approx0.7$ at $\theta_\text{in}=\pm\SI{90}{\degree}$. The approximately 50-50 odds of exiting dorsally or ventrally at $\theta_\text{in}=0$ is consistent with the response being determined by the sensed repellent gradient in the D-V direction upon encountering the boundary (\textit{i.e.}~the gradient magnitude in the direction perpendicular to $\theta_\text{in}$, $|\nabla C_\perp|=|\nabla C|\sin\theta_\text{in}$, where $|\nabla C|$ is the gradient magnitude), which has zero magnitude at $\theta_\text{in}=\SI{0}{\degree}$. Furthermore, the shape of the symmetric response profile in $P_\text{escape}^\text{anat}(\theta_\text{in})$ is consistent with a response modulation proportional to $|\nabla C_\perp|$. Assuming that the response in  $P_\text{escape}^\text{anat}$ is some function of that transverse gradient ($P_\text{escape}^\text{anat}=f(|\nabla C_\perp|)$), and further that $|\nabla C_\perp|$ is sufficiently small, it is then reasonable to expand $f(|\nabla C_\perp|)$ to linear order in $\sin\theta_\text{in}$ to obtain $P_\text{escape}^\text{anat}\approx f(0)+A\sin{\theta_\text{in}}$, where $f(0)=0.5$ and $A$ is a constant. This function obtains a good fit to the observed $P_\text{escape}^\text{anat}$ profile as a function of $\theta_\text{in}$ (Figure \ref{fig:figure4}H, dashed curve) with  $A\approx0.2$. Thus, the observed symmetric profile of the anatomical-frame escape probability is compatible with a simple model that makes just two assumptions (i) the worm upon encountering the repellent boundary faces a binary decision: whether to exit the escape turn dorsally or ventrally, and (ii) the probability $P_\text{escape}^\text{anat}$ of making the "correct" decision (\textit{i.e.} exiting in the D-V direction opposite to that of encounter) is limited by the magnitude of the sensed gradient in the D-V direction upon encountering the boundary at an angle $\theta_\text{in}$.

\section{Discussion}
By developing a novel behavioural assay that enables tracking multiple \textit{C. elegans} individuals over long times ($\SI{2}{\hour}$), we quantified the statistics of turning behaviour in both exploration and escape navigation contexts. The data revealed significant biases in both gradual- and sharp-turn behaviors, which impose constraints on exploration and escape performance, respectively. In the context of exploration, quantifying the diversity of motility phenotypes within an isogenic population allowed us to identify the subset of reorientation behaviours that correlate most strongly with exploration performance, from which we derived a minimal model of motile trajectories. Analysis of this model identified a novel optimality principle for maximising exploratory propensity under the constraint of finite gradual-turn bias.
Similarly, in the context of escape navigation, studying the statistics of sharp turn directions as a function of the encounter angle with the repellent gradient revealed how worms optimize their behavior despite constraints likely arising from biases  in control physiology and from limited information about its orientation with respect to the repellent gradient.

\subsection{Optimizing exploratory propensity under gradual-turn bias requires nonzero angular diffusivity}
Optimality is useful as a guiding concept for studying biological design, given that natural selection tends to drive some performance measures of the system towards a maximum. In the context of behavior, identifying the relevant performance measure being optimised provides a framework to study the design of the underlying control strategies and their physiological implementation. However, identifying these objective functions on which selection acts is often not trivial, as in nature, one can expect selection to be acting simultaneously on multiple such performance criteria that may be in conflict and, as a result, impose constraints on one another. Within our study, we identified as a performance measure for exploration the trajectory persistence length, and found that maximizing this performance under the constraint of finite gradual-turn bias (\textit{i.e.}~trajectory curvature) requires a nonzero rotational diffusivity.
A gradual-turn bias resulting from a lack of control of orientation is not unique to the worm, but has been observed across a wide variety of organisms, for navigational tasks in environments that lack sufficient sensory cues for direction. The causal factors leading to such biases are difficult to resolve, and could be manifold. For example, loopy trajectories created by blindfolded humans have been hypothesised to arise from anatomical asymmetries \cite{Maus2014}, an imperfect 'sense of straight ahead' \cite{Bestaven2012} and accumulating noise in the sensimotor system \cite{Souman2009}.
Recent research on \textit{Drosophila} revealed a persistent crawling preference which is not linked to body asymmetries \cite{Ayroles2015}, but suggests a neuronal origin \cite{Buchanan2015}. \textit{C. elegans} can be used as a useful model organism to fundamentally study the (neuronal) origin of such a rotational drift. Our minimal model of exploratory trajectories (Eq.~\ref{eq:persLength}) does not address the causal factors leading to such gradual-turn biases, but rather predicts their consequences. In particular, for any given trajectory curvature $\kappa$, the model reveals that the maximal exploratory propensity (\textit{i.e.}~ persistence length $P$) will be achieved when the rate of random orientation $\epsilon$ (which in \textit{C. elegans} is dominated by orientational diffusion) exactly balances the magnitude of $\kappa$, and our data indicate that \textit{C. elegans} trajectories indeed demonstrate, on average, this optimal balance.
Thus, although the trajectory curvature induced by the worms' gradual-turn bias does impair exploratory performance, the latter achieves the greatest value possible under that constraint of finite bias to enhance the spatial extent of exploration.

Interestingly, however, we found considerable variability at the level of individual worms in both gradual-turn bias $\kappa$ and random reorientation $\epsilon$, with only a weak correlation between these parameters across individuals (Figure~\ref{fig:figure4}D). Thus, although this optimal balance between $\kappa$ and $\epsilon$ is evidently achieved at the level of the population average, this balance is not tightly controlled at the level of individuals.
From a mechanistic point of view, it is in fact interesting that these variations in gradual-turn bias and angular diffusion appear to be nearly independent of one another, given that they are both represent errors and/or fluctuations in the body-wave dynamics driving the worms' undulatory propulsion. Naively, one might expect a stronger correlation between these, if they were both the product of finite control over the locomotor wave.
One compelling hypothesis is that the angular diffusivity due to finite errors in locomotor wave control is in fact considerably lower and that the observed diffusivity is actually dominated by yet another type of reorientation behavior. It has been documented by Kim et al. \cite{Kim2011} that \textit{C. elegans} trajectories can feature a high frequency of "shallow turns" during runs with reorientation angles much $<\SI{90}{\degree}$. In our analysis, reorientations due to such shallow turns would not be detected as sharp turns, but instead contribute to the magnitude of the angular diffusivity. Interestingly, inspection of reorientation statistics between consecutive body waves in our data revealed a broad distribution that could be fitted by the sum of two Gaussians
(Figure~\ref{fig:figure3}--Figure Supplement~2). Thus, it is conceivable that the broader of these two Gaussians, which contributes the majority of the variance ($\sigma=\SI{28.3}{\degree}\pm \SI{0.4}{\degree}$, as compared to $\sigma=\SI{13.5}{\degree}\pm\SI{0.1}{\degree}$ for the narrower Gaussian), reflects shallow turns.
Finally, regardless of the underlying mechanisms, the nearly uncorrelated variation in gradual-turn bias and orientational diffusion leads to large variation in exploratory performance, which may be interpreted as a bet-hedging strategy \cite{Slatkin1974,Philippi1989}. Exploring new regions of space and exploiting local resources is a well-known trade-off in foraging strategies, and hence expressing a diversity of phenotypes along this exploration-exploitation axis could provide isogenic worm populations with an effective adaptive strategy in rapidly changing and/or information-scarce environments where sensory modulation of behaviour is less effective \cite{Xue2019}. 

Our minimal model of trajectory statistics combines the effects of sharp turns, orientational diffusion, and gradual-turn bias, and accurately predicts the experimentally observed trajectory persistence lengths. The simplicity of the model offers key insights into how variations in these parameters interdependently affect this measure of exploration performance, and provides a basis for future studies that examine the effect of perturbations such as genetic mutations or neural ablations. Furthermore, its construction is sufficiently general that it can be readily applied to any organism (or motile particle/agent) whose motion can be described by trajectory curvature, effective diffusivity, and intermittent sharp turns, for instance the run-and-tumble motion of swimming bacteria near surfaces that induce curved runs \cite{Lauga2006}, or other nematodes such as larvae \textit{Ancylostoma tubaeforme} \cite{Croll1975_2}.
Our model reveals that any nonzero trajectory curvature sets an upper limit to the persistence length, which in the absence of external guiding cues significantly reduces the exploration performance.
Whether and to what extent \textit{C. elegans}' gradual-turn bias also impacts the performance in other environments (e.g. during chemotaxis) remains an open question and will provide fertile ground for future studies.
During taxis strategies, gradual-turn bias might be expected to reduce, but not abolish, taxis efficiency. The head-bend mutant \textit{unc-23} creates spiral-shaped tracks with a stronger curvature compared to worms in this study, but can still perform navigational tasks like chemotaxis \cite{Waterston1980,Ward1973,PierceShimomura2005}. Of particular interest would be to investigate the relationship between the gradual-turn bias we have observed here in the absence of environmental gradients and the 'weathervaning' (klinotaxis) strategy of chemotaxis that has been shown to steer trajectory curvature in response to strong environmental gradients \cite{Iino2009}. For example, testing whether the weathervaning response completely overrides the gradual-turn bias, or acts additively could shed light on whether the bias and weathervaning response are controlled by the same neural circuitry.

\subsection{Context-dependent modulation of sharp-turn statistics symmetrizes escape performance under biasing constraints}
The worms' sharp-turn response to an acute aversive stimulus provides another example of constrained optimization of behavior. We examined escape from a strong chemorepellent (SDS) that worms encountered as a spatial gradient as they approached the arena's boundary, and assessed escape performance as a function of the angle of encounter with the repellent boundary. By analyzing thousands of such escape-turn events, we found that control of escape direction is achieved by selecting from a discrete repertoire of posturally distinct sharp-turn types ($V\Omega$,$V\delta$,$D\Omega$), rather than continuous modulation of sharp-turn amplitudes. 

Although the exact performance measure being optimized by this decision is unknown, it is natural to expect that the goal of the escape response is to reorient the worm's movement away from the repellent source. We therefore considered as the performance measure the probability $P_\text{escape}$ that the escape turn successfully reorients the worm away from the repellent source, defined in two contrasting ways. The first, $P_\text{escape}^\text{env}$, defined in the reference frame of the environment, corresponds to the probability that worms exited escape turns in the direction down the repellent gradient. The second, $P_\text{escape}^\text{anat}$, defined in the reference frame of the worm's own anatomy, corresponds to the probability that the exit angle of the worm pointed away from the side of the body that faced the repellent boundary upon encounter. Interestingly, $P_\text{escape}^\text{env}$ was asymmetric with respect to the angle of encounter with the repellent boundary when performance was defined in the reference frame of the environment, whereas $P_\text{escape}^\text{anat}$ was symmetric with respect to the encounter angle.

Asymmetry in the environmental-frame performance $P_\text{escape}^\text{env}$ might reflect biases in underlying sharp-turn behavior, which in turn could be due to  anatomical or physiological constraints.
Even when the aversive cue was on the ventral side, dorsal turns were less frequent than ventral turns. Furthermore, dorsal $\Omega$-turns were generally more shallow than ventral $\Omega$-turns and there were no dorsal $\delta$-turns. While this asymmetry between dorsal and ventral turns is largely consistent to that observed in the freely crawling context, it is striking that it persists even in the escape context involving a strongly aversive SDS stimulus. Hence, the asymmetric sharp-turn statistics might indicate a fundamental biasing constraint that renders dorsal turns anatomically or physiologically less favorable.

Despite this strong biasing constraint, however, the repertoire of three turn types ($V\Omega$,$V\delta$,$D\Omega$) is sufficiently rich that, even with an overall preference for ventral turns, it should in principle be possible to achieve the theoretically optimal escape response $P_\text{escape}^\text{env}=1$ at any encounter angle. The lower values of $P_\text{escape}^\text{env}$ we observed thus suggests that escape performance might be limited by a finite accuracy in perceiving the gradient encounter angle $\theta_\text{in}$. Accurate evaluation of $\theta_\text{in}$ would require precise sensory measurements of spatial differences in the SDS concentration within the anatomical reference frame, which could be challenging at the length scale of the worm and within the time scale of an escape response.  But even if we accept that finite sensory information ultimately limits escape perfmance (such that $P_\text{escape}^\text{env}<1$), how can we explain the observed asymmetry in $P_\text{escape}^\text{env}$ as a function of the encounter angle $\theta_\text{in}$. There is no systematic dorsal-ventral asymmetry in the sensory neurons \cite{Hilliard2002} and the chemorepellent constitutes an equal threat irrespective of whether it is encountered on the dorsal or ventral side of the body. While we cannot rule out that individual worms have some dorsal-ventral bias in how they perceive the cue, there is no obvious reason for a perceptual bias across the population of worms we sampled here. We are thus inclined to consider that the anatomical-frame performance measure $P_\text{escape}^\text{anat}$, with its more symmetric performance profile, may well approximate the objective function being optimized by neural computations in \textit{C.~elegans}. According to this performance measure, successful escape requires only a determination of whether the repellent threat is on the dorsal or ventral side of the worm's anatomy, and then executing a turn that results in an exit on the opposite side of the body. Basing the escape behavior on this binary decision task (captured by the $P_\text{escape}^\text{anat}$ performance measure) would have the advantage that it does not require a precise determination of the gradient angle, which is a non-trivial task, especially if non-deleterious escape depends on a quick response at low concentration levels of the chemorepellent.

Interestingly, while we observed that the escape response to the SDS chemorepellent used here deployed the full repertoire of sharp-turn types ($V\Omega$,$V\delta$,$D\Omega$), previous studies have found that worms exclusively execute turns of type $V\Omega$ during escape \cite{Broekmans2016, Donnelly2013, florman2022co}. We suggest that this difference is likely due to the contrasting aversive stimuli used to trigger escape responses in those studies. Whereas in our study, the escape was triggered by encounter with a spatial gradient of a chemorepellent, in those studies escape was triggered by a laser-induced heat pulse \cite{Broekmans2016} or touch with an eyelash \cite{Donnelly2013, florman2022co}. In contrast to stimuli encountered as gradients, the latter more impulsive stimuli likely do not carry directional information, and it is possible that the worm defaults to the "preferred" turn type, namely $V\Omega$, in lieu of any spatial cues that would favor an alternative behavioral decision.

\subsection{Neural control of reorientation statistics: possible targets for future studies}
The control of reorientation statistics (or lack thereof) described in this study, raises the question which neuronal circuitry generates these behavioral patterns.
The neuronal signalling network underlying weathervaning-type control of \textit{C. elegans} trajectory curvature \cite{Iino2009} has yet to be uncovered, but it possibly involves sampling the environment by oscillatory-like head-swings required for propulsion \cite{Izquierdo2010,Kato2014}, possibly through the SMB neuron associated with head oscillations and gradual turning via interneurons like AIY and AIZ \cite{Izquierdo2015}. Killing of the SMB neurons results in large head-swings during forward crawling with high-curvature trajectories \cite{Gray2005}, suggesting a possible mode of control for the crawling bias.
However, it has to be seen to what extent the set of neurons involved in control during escape and weathervaning responses overlap, given that weathervaning is described at the level of more gradual rather than sharp reorientations \cite{Iino2009} and the SDS sensing ASH neuron directly synapses to the reversal triggering AVA neuron, bypassing much of the chemotaxis circuitry (although ASH is connected to AIA \cite{Murayama2013}).

Some of the neurons reported to affect sharp turn behaviors (that might be involved in control during escape) are RIM, RIV, RIB, and SMD. RIV (and RIB) activity rises at the onset of the turn \cite{Wang2020} and RIV ablated worms lack a ventral bias \cite{Gray2005}, suggesting their involvement in modulation of $P(D)$. RIM is a tyraminergic neuron that aides in the hyperpolarization of ventral muscles to execute the ventral $\Omega$-turn \cite{Donnelly2013,Kagawa2018}, and therefore might be related to the $\delta$-turn modulation. Similarly, ablation of the SMD neuron results in sharp turns with smaller reorientation angles \cite{Gray2005} which suggests its involvement in the sharp turn amplitude. The SMDD, and also DVA neurons have been reported to provide proprioceptive feedback (\textit{i.e.}~sensing of the body bending angle) \cite{Li2006,Yeon2018}, and therefore might be involved in the regulation of (sharp) turn direction, enabling angular-dependent control.
In addition, asymmetric feedback from such proprioceptive neurons or mechanosensory neurons (worms lacking mechanosensory neurons, PVD and FLP produce circular trajectories \cite{Cohen2012}) might be the cause of a rotational bias. It would be exciting in further studies to examine the effects of perturbing these neurons (via e.g. ablation, genetic mutations, or optogenetic stimulation) on the control of reorientation behaviours we studied here. Finally, our measurement and analysis of escape-turn statistics suggests that a discrete, rather than continuous, decision process might underlie the control of sharp turns during the escape response. Interestingly, a previous study that focused on random search behavior also yieled a discrete state model for motor command units  in \textit{C. elegans} \cite{roberts2016stochastic}. Whether such discrete decision processes are implemented by individual neurons, or in the collective activity patterns of multiple neurons (e.g. as neural circuits) would be exciting to explore in future experiments that combine behavioral and neural measurements.

\subsection*{Conclusions}
Our results revealed how \textit{C.~elegans} reorientation statistics demonstrate significant biases. In the context of random spatial exploration, the reorientation parameters appear to be tuned to maximise exploratory propensity, under the constraint of finite bias. 
In the context of escape, worms demonstrated the ability to strongly modulate escape performance symmetrically in both dorsal and ventral directions, despite a strong underlying sharp-turn bias in the ventral direction. Our minimal model hints at a binary control logic of escape behavior and provides a basis for further investigation of the relationship between reorientation behaviors, their mechanistic origins, and their functional consequences.

\section{Methods and Materials}

\paragraph{Behavioural experiments}
Worms are cultivated on NGM plates containing HB101. A copping ring (a $\SI{38}{\milli\meter} \times \SI{38}{\milli\meter}$ square with rounded corners and a total surface area of $\SI{13.8}{\milli\meter}^2$) soaked in $1\%$ SDS is put on a food-free NGM plate. $8$ well fed young adult worms are washed for $\SI{15}{\minute}$  in a $\SI{1}{\micro\liter}$ M$9$ solution and pipetted onto the arena. The motility of the plate is recorded for $\SI{2}{\hour}$ with $\SI{11.5}{\hertz}$ using an PointGrey GX-FW-60S6M-C. During the experiment, the average speed of the worms remained mostly constant (Figure\ref{fig:TurnPartitioning}--Figure Supplement~2).

\paragraph{Tracking of worms}
Worm are tracked using custom tracking software written in python. Collisions are resolved semi-automatically, using a combination of the worm lengths, collision duration and direction of motion before and after the collision of each worm. Ambiguous collisions, or collisions involving more than $2$ worms are resolved by hand. Automatically resolved collisions are all manually inspected.

\paragraph{Sharp turn extraction}
Sharp turns are differentiated from gradual turns by the body posture's topology. During a sharp turn, the worm folds onto itself creating a doughnut topology. 
These postures cannot trivially be extracted from the spline of the binarized worm image. Therefore, a customized version of a previously published algorithm by Broekmans et. al. is used to solve sharp turn body postures \cite{Broekmans2016}, using the OIST’s Sango cluster parallel computation cluster.
A small fraction of turns were not successfully resolved by the algorithm. Unsuccessful turns were identified using the algorithm's image-comparison metric which provides an error value for each frame in the turn. Turns with an error > 12 for at least 3 consecutive frames were flagged as unsuccessful and not included in the analysis.
$10625$ out of $12475$ sharp turns could be resolved using this method. A manual annotation of a random selection of turns that could not be resolved did not reveal obvious biases towards a certain turn type.

Individual sharp turn events correspond to intervals, in which the worm body continuously forms a doughnut topology (gaps $<\SI{1}{s}$ are allowed). 
The sharp turns are associated with large absolute values of the third Eigenworm $|A_3|$ \cite{Stephens2008} and we identify the bending peak as the maximum in $|A_3|$ for each sharp turn. Finally, we determine the start and end point of each turn as the closest extrema around this peak. If $A_3>0$ at the peak, the start and end points are given by the closest minima with $A_3<5$. If $A_3<0$ at the peak, we identify the closest maxima with $A_3>-5$.

\paragraph{Distinguishing sharp turn types}
We distinguish "spontaneous" turns, which occur uniformly in the whole domain and thus without an apparent stimulus, from escape responses, which predominantly occur close to the repellent SDS boundary. 
An escape response is defined as a sharp turn following a reversal. A spontaneous turn is occurring without a preceding reversal. Furthermore, we exclude all sharp turns, where a reversal is immediately following the turn. We define a reversal as a backward motion for more than half of the time 
within an interval of $\SI{2}{\second}$ (\textit{i.e.}~$23$ frames). From the $10625$ resolved sharp turns, we obtain: $5799$ spontaneous turns, $3866$ escape turns, and $960$ excluded turns (which either are followed by a reversal or where it is undetermined whether there is a reversal). For Figure~\ref{fig:figure4}, we focus on the $2943$ escape turns close to the boundary (\textit{i.e.}~less than $\SI{7}{\mm}$ away).

\paragraph{Ventral annotation}
The resolution of the camera did not allow to visually separate dorsal from ventral. However, we find that for each worm $\delta$ turns almost exclusively in a consistent direction. We infer this to be the ventral direction.

\paragraph{Testing worm-to-worm variability by Monte-Carlo sampling}
To test if the observed variability is not trivially explained by stochasticity, measured worm-to-worm variability in turning behavior is compared with simulations from a stochastic model. In the model all worms are assumed to follow the same stochastic process determined by population average statistics and each turn is independent.
The sharp turn rate stochastic process is described by the population average interval distribution (Figure~\ref{fig:figure1}--Figure Supplement~1). 
Drawing from this distribution, using Monte Carlo sampling, a simulated number of sharp turns can be obtained for the duration of each worm in the measurement, resulting in a different distribution of turn frequencies (Figure~\ref{fig:figure1}B).
Since the simulations are subject to stochasticity as well, the process is repeated $1000$ times. The measured distribution was found to consistently posses larger variation compared to the simulated distributions (Figure~\ref{fig:figure1}--Figure Supplement~2 A).
In the case of worm-to-worm variability in $P(D)$ and $P(\delta|V)$, the corresponding stochastic process is a coin flip (\textit{i.e.}~Bernoulli process), using the population average probability. A number of dorsal or $\delta$ turns is randomly drawn from a binomial distribution using the population average probability and a number of coin flips specified for each worm by the number of sharp turns in the case of the $P(D)$ and the number of ventral sharp turns in the case of $P(\delta|V)$ (Figure~\ref{fig:figure1}B,C). Likewise, the resulting distributions were found to be more variable compared to the simulated distribution for a significant fraction of the simulations (Figure S\ref{fig:figure1}--Figure Supplement~2 B,C). Variation in sharp turning frequency can be partially attributed to batch effects of worms measured simultaneously on the same plate. However, a larger fraction of the observed variability in both frequency, $P(D)$ and $P(\delta|V)$ is of unknown origin and might stem likewise from experiences of the environment as well as from intrinsic stochasticity (Figure S\ref{fig:figure1}--Figure Supplement~3).\\

\paragraph{Extraction of model parameters}
To compute the curvature and rotational diffusion, first the unwrapped (accumulative) average worm body orientation in a window of $\SI{5}{\second}$ around sharp turn events and worm collisions is excluded. We found that the average body orientation is an accurate proxy of the worm's body velocity bearing during runs. However, it is well-defined throughout the trajectory, even at low speeds, and thus the accumulative angle does not suffer from artifacts.

The orientation is computed as a function of trajectory length, by evaluating it at equally spaced intervals of $\SI{100}{\micro\meter}$. 
The curvature is estimated in windows of $\SI{15}{\minute}$ as the average spatial rotation rate.
The rotational diffusion $D_{\psi}$ is extracted by fitting the function $y=2D_{\psi}x+bx^2$ to the MSD (evaluated up to $\SI{1}{\milli\meter}$) of the spatial orientation.

\paragraph{Computing the persistence length}
The persistence length is extracted using the MSD of the worm's centroid position (evaluated at the $\SI{100}{\micro\meter}$ intervals along the trajectory contour). For a diffusive process with (translational) diffusion coefficient $D_{\text{t}}$ in $n$ dimensions, the mean-squared displacement as a function of time $t$ follows the equation $\text{MSD}=2nD_{\text{t}}t$. Assuming a constant speed 
of diffusive process $s=x/t$, $2$ dimensions, and a definition $P\equiv D_{\text{t}}/s$, we obtain that $P=\text{MSD} / 4x$. Theoretically, in the diffusive regime a constant can be fitted to $MSD/x$. Due to confinement this curve decreases when the MSD approaches the size of the arena. Furthermore, $\text{MSD}/x$ fluctuates slightly, because trajectories are described by a random process. Therefore, the persistence length is evaluated as the average value where the MSD is linear; in the range where the slope of $\log \text{MSD}$ vs $\log x$ is in between $0.9$ and $1.1$.

\paragraph{Computing the gradual turn bias decorrelation time $\tau$.}
The autocorrelation function (ACF) cannot be accurately estimated on a single-worm basis, because the fluctuation time scales of $\kappa$ is of the same order as the length of the measurement. It can be computed on a population level, assuming that each worm follows the same stochastic process.
With this assumption we compute the ACF using the population variance and mean ($0$) of $\kappa$ for each worm and average across worms. The resulting curve is fitted to the function $\langle\text{ACF}_\kappa(t)\rangle= e^{-t/\tau}$ using the 'curve\_fit' function of the scipy python package. To obtain a confidence interval, the process of computing the ACF and fitting $\tau$ is repeated $1000\times$ after bootstrapping for worms. The reported error is the standard deviation of bootstrapped values of $\tau$.

\section{Acknowledgments}


We thank Laetitia Hebert, and members of the Shimizu and Stephens groups, as well as Dr. Andre Brown and his lab members, for helpful discussions. We are grateful for the help and support provided by the Scientific Computing section of the Research Support Division at OIST.

\bibliography{elife-template}


\appendix
\begin{appendixbox}
\label{first:app}
\section{Derivation of the model}
We consider a simple random walk as a minimal model for {\it C.~elegans} locomotion. 
We take three observation about the worm movement into account: (i) Worms move along a curved trajectory, for which we assume a constant average curvature $\kappa$ [$\si{\milli\meter}^{-1}$]. (ii)  This run is interrupted by occasional random reorientation events of rate $\zeta/s$ [$\si{\milli\meter}^{-1}$] with the (constant) movement speed $s$ and the reorientation frequency $\zeta$. (iii) The worm trajectory is subject to rotational diffusion with diffusion coefficient $D_{\psi}$ [$\si{\milli\meter}^{-1}$]. We neglect reversals, such that the speed $s$ is always positive.

First, we neglect the rotational diffusion and only consider random orientation events and a curved trajectory.
The probability density to turn again after moving a length $x$ since the last turn is
\begin{equation}
p(x)=(\zeta/s)\,e^{-x\zeta/s}\,.
\end{equation}
As the worm moves along the perimeter of circle, the Euclidean distance between turns is
\begin{equation}
r=(2/\kappa)\,\sin(x\kappa/2)\,.
\end{equation}
We can map this random walk to a wait-and-jump process, where the worm waits for a time $t$ at a position and jumps a distance $r$. As a consequence, the effective (translational) diffusion coefficient in $d=2$ dimensions is
\begin{equation}
D_{\text{t}}=\langle r^2\rangle/(2d\langle t\rangle)=\langle r^2\rangle/(4\langle t\rangle)\,.
\end{equation}
In the following, we will consider the persistence length $P=D_{\text{t}}/s$:
\begin{equation}
P=\langle r^2\rangle/(4\langle x\rangle)\,,
\end{equation}
where $x=st$ with a constant speed $s$.
We can derive the average run length:
\begin{equation}
\langle x\rangle=\int_0^{\infty}x\,p(x)\,dx=\int_0^{\infty}x\,e^{x\zeta/s}\zeta/s\,dx=1/(\zeta/s)\,.
\end{equation}
The average distance from the origin is
\begin{align}
&\langle r^2\rangle=\int_0^{\infty}\Big((2/\kappa)\,\sin(x\kappa/2)\Big)^2\,e^{-x\zeta/s}\zeta/s\,dx=2/(\kappa^2+(\zeta/s)^2)\\
&P=\frac{\zeta/s}{2\kappa^2+2(\zeta/s)^2}\,.\label{eq:Pfirst}
\end{align}
The same result has been derived by Martens {\it et al.}~in analogy to electrons in a magnetic field \cite{Martens2012}.

We can consider two limiting cases.
If the worm turns very often such that $\kappa\gg\zeta/s$ and the path is straight between reorientation events, the persistence length decreases with higher turn frequency according to $P\propto 1/\zeta$. In contrast, for circular trajectory with $\kappa\ll\zeta/s$, reorientation events are beneficial to explore a larger area and thus $P\propto \zeta$.

Next, we include rotational diffusion. If the runs have no gradual turning bias ($\kappa=0$), the effective diffusion coefficient on large scales is \cite{Cates2013}
\begin{equation}
P=1/(2\epsilon)\,,
\end{equation}
with an effective reorientation frequency in $d=2$ dimensions
\begin{equation}
\epsilon=(d-1)D_{\psi}+\zeta/s=D_{\psi}+\zeta/s\,.
\end{equation}

\end{appendixbox}
\addtocounter{appendix}{-1}
\begin{appendixbox}

The rotational diffusion has an analogous effect as abrupt reorientation events on sufficiently large scales.
In this spirit, we replace $\zeta/s$ by $\epsilon$ in Eq.~\ref{eq:Pfirst} and obtain
\begin{align}
&P=\epsilon/(2\kappa^2+2\epsilon^2)\,.\label{eqfinalres}
\end{align}
This solution agrees very closely with simulations over a wide range of parameters (Figure~\ref{fig:figure3}--Figure Supplement~1).
In the case that the sharp turn does not fully randomize the reorientation, but is biased along the direction of motion with $\alpha=1- \langle \cos \Delta \theta \rangle$, $\zeta$ has to be replaced by $\bar{\zeta}=\zeta \alpha$ \cite{Taktikos2013,Locsei2007}.

\end{appendixbox}
\clearpage

\end{document}